\documentclass[10pt,twocolumn,letterpaper]{article}

\usepackage{xspace}
\newcommand{\method}{C3\xspace}

\usepackage[]{cvpr}      %

\usepackage[dvipsnames, table]{xcolor}

\usepackage{microtype}

\usepackage[subpreambles=true]{standalone}

\usepackage{tikz}
\usepackage{pgfplots}
\usepackage{pgfplotstable}
\usetikzlibrary{calc}
\usetikzlibrary{positioning}
\usetikzlibrary{angles,quotes}
\usetikzlibrary{backgrounds}
\usetikzlibrary{external}
\usetikzlibrary{fit}
\usetikzlibrary{arrows}
\usetikzlibrary{arrows.meta}
\usetikzlibrary{shapes.symbols}
\usetikzlibrary{shadings}
\usetikzlibrary{shapes}
\usetikzlibrary{spy}
\usetikzlibrary{fadings}
\usetikzlibrary{bayesnet}
\usetikzlibrary{matrix}
\usetikzlibrary{decorations.pathmorphing, decorations.pathreplacing, patterns,calligraphy}
\usetikzlibrary{3d}
\usepgfplotslibrary{groupplots}
\pgfplotsset{compat=1.18}

\usepackage[normalem]{ulem}

\usepackage{colortbl}
\usepackage{pdflscape}
\usepackage{listings}
\usepackage{multirow}
\newcommand{\midrulegray}{\arrayrulecolor{lightgray}\midrule\arrayrulecolor{black}}

\newcommand{\jax}{\texttt{JAX}\xspace}

\newcommand{\fvcore}{\texttt{fvcore}\xspace}
\newcommand{\pytorch}{\texttt{PyTorch}\xspace}
\newcommand{\deepspeed}{\texttt{DeepSpeed}\xspace}
\newcommand{\ptflops}{\texttt{ptflops}\xspace}

\newcommand{\enticon}[1]{\tikz[baseline=-0.7ex]{\node[draw, shape=rounded rectangle] {\sffamily E\,#1}}}
\newcommand{\synthicon}[1]{\tikz[baseline=-0.7ex]{\node[draw, shape=rounded rectangle] {\sffamily S\,#1}}}

\usepackage{pifont}%
\newcommand{\cmark}{\ding{51}}%
\newcommand{\xmark}{\ding{55}}%
\usepackage{fontawesome}

\usepackage{array}
    \newcolumntype{C}{>{$}c<{$}}
    \newcolumntype{L}{>{$}l<{$}}
    \newcolumntype{R}{>{$}r<{$}}
    \newcolumntype{M}{>{\centering\arraybackslash$}m{1.4cm}<{$}}

\newcommand{\vz}{\ensuremath{\mathbf{z}}\xspace}
\newcommand{\vx}{\ensuremath{\mathbf{x}}\xspace}
\newcommand{\vu}{\ensuremath{\mathbf{u}}\xspace}

\newcommand{\synthnet}{\ensuremath{f_\theta}\xspace}
\newcommand{\entropynet}{\ensuremath{g_\psi}\xspace}

\newcommand{\cc}{COOL-CHIC\xspace}
\newcommand{\cctwo}{COOL-CHICv2\xspace}

\newcommand\blfootnote[1]{%
  \begingroup
  \renewcommand\thefootnote{}\footnote{#1}%
  \addtocounter{footnote}{-1}%
  \endgroup
}

\definecolor{C0}{HTML}{1F77B4}
\definecolor{C1}{HTML}{FF7F0E}
\definecolor{C2}{HTML}{2CA02C}
\definecolor{C3}{HTML}{D62728}
\definecolor{C4}{HTML}{9467BD}
\definecolor{C5}{HTML}{8C564B}
\definecolor{C6}{HTML}{E377C2}
\definecolor{C7}{HTML}{7f7f7f}
\definecolor{C8}{HTML}{bcbd22}
\definecolor{C9}{HTML}{17becf}
\colorlet{C0light}{C0!70!white}
\colorlet{C1light}{C1!70!white}
\colorlet{C2light}{C2!70!white}
\colorlet{C3light}{C3!70!white}
\colorlet{C4light}{C4!70!white}
\colorlet{C0vlight}{C0!20!white}
\colorlet{C1vlight}{C1!20!white}
\colorlet{C2vlight}{C2!20!white}
\colorlet{C3vlight}{C3!20!white}
\colorlet{C4vlight}{C4!20!white}
\colorlet{C0dark}{C0!70!black}

\colorlet{verylightgray}{lightgray!40!white}

\newcommand{\Beta}[2]{\ensuremath{\mathrm{Beta}(#1, #2)}}

\makeatletter
\renewcommand\paragraph{\@startsection{paragraph}{4}{\z@}%
                                    {1.25ex \@plus1ex \@minus.2ex}%
                                    {-1em}%
                                    {\normalfont\normalsize\bfseries}}
\makeatother

\definecolor{cvprblue}{rgb}{0.21,0.49,0.74}
\usepackage[pagebackref,breaklinks,colorlinks,citecolor=cvprblue]{hyperref}

\def\paperID{15913} %
\def\confName{CVPR}
\def\confYear{2024}

\title{C3: High-performance and low-complexity neural compression\\ from a single image or video}
\author{Hyunjik Kim*, Matthias Bauer*, Lucas Theis, Jonathan Richard Schwarz, Emilien Dupont*\\
	Google DeepMind\\[0.5ex]
	{\normalsize *Equal contribution \qquad }\\
{\normalsize Corresponding authors: \tt \{hyunjikk, msbauer, edupont\}@google.com}
}

\begin{document}
\maketitle

\begin{abstract}
Most neural compression models are trained on large datasets of images or videos in order to generalize to unseen data. 
Such generalization typically requires large and expressive architectures with a high decoding complexity.
Here we introduce \method, a neural compression method with strong rate-distortion (RD) performance that instead overfits a small model to each image or video separately.
The resulting decoding complexity of \method can be an order of magnitude lower than neural baselines with similar RD performance.
\method builds on COOL-CHIC (\citet{ladune2023cool}) and makes several simple and effective improvements for images. We further develop new methodology to apply \method to videos.
On the CLIC2020 image benchmark, we match the RD performance of VTM, the reference implementation of the H.266 codec, with less than 3k MACs/pixel for decoding.
On the UVG video benchmark, we match the RD performance of the Video Compression Transformer (\citet{mentzer2022vct}), a well-established neural video codec, with less than 5k MACs/pixel for decoding.
\end{abstract}
    
\section{Introduction}
\label{sec:intro}

Most neural compression models are based on autoencoders \cite{balle2016end, theis2017lossy}, with an encoder mapping an image to a quantized latent vector and a decoder mapping the latent vector back to an approximate reconstruction of the image. \blfootnote{JRS is now at Harvard University. \\ Detailed author contributions are at the end of the paper.}
To be practically useful as codecs, these models must \textit{generalize}, \ie, the decoder should be able to approximately reconstruct any natural image. Such a decoding function is likely to be complex and expensive to compute.
Indeed, while most neural codecs enjoy very strong rate-distortion (RD) performance \cite{cheng2020learned, he2022elic, jiang2022multi}, their decoding complexity can make them impractical for many use cases, particularly when hardware is constrained, \eg, on mobile devices \cite{le2022mobilecodec, van2023mobilenvc}. As a result, designing low complexity codecs that offer strong RD performance is one of the major open problems in neural compression \cite{yang2023introduction}. %

\begin{figure}[t]
  \centering
  \includestandalone[mode=image, width=\columnwidth]{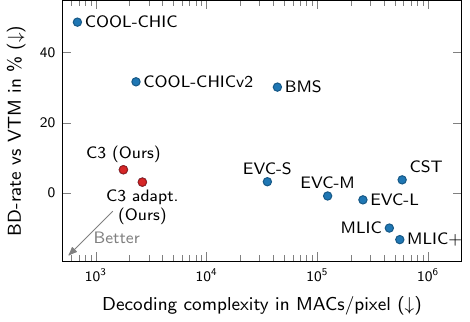}
    \caption{Rate-distortion performance (BD-rate) vs. decoding complexity on the Kodak image benchmark. Our method, \method, achieves a better trade-off than existing neural codecs.}
   \label{fig:bd-rate-vs-macs}
\end{figure}

\begin{figure*}
\centering
\includestandalone[width=\textwidth, mode=image]{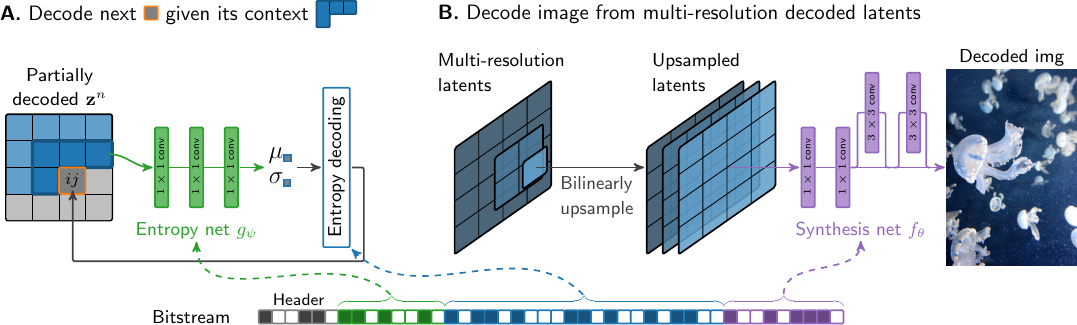}
\caption{Decoding the bitstream into an image with \cc and \method. \textbf{\sffamily A.} A latent entry $\widehat\vz^n_{ij}$ (\protect\tikz[scale=0.2, baseline=0.1ex]{\protect\draw[step=1,draw=C1,fill=gray, thick, rounded corners=0.5] (0, 0) rectangle (1, 1);}) is autoregressively decoded by applying the entropy network $\entropynet$ to the context $\mathrm{context}(\vz^n; (i, j))$
(\protect\tikz[scale=0.175, baseline=.75ex]{
	\protect\draw[step=1,fill=C0, draw=none] (1, 1) rectangle (4, 2);
	\protect\draw[step=1,fill=C0] (1, 1) grid (4, 2);
	\protect\draw[step=1,fill=C0, draw=none] (1, 0) grid (2, 1) rectangle (1, 0);
	\protect\draw[C0dark, thick, rounded corners=0.5] (1, 0) -- (2, 0) -- (2, 1) -- (4, 1) -- (4, 2) -- (1, 2) -- cycle;
}).
\textbf{\sffamily B.} The decoded latent grids at multiple resolutions are first upsampled and then decoded into image space using the synthesis network $\synthnet$. Figure adapted from \citet{leguay2023low}.}
\label{fig:cc_model_overview}
\end{figure*}

Recently, an alternative approach to neural compression called COIN was proposed \cite{dupont2021coin}. Instead of generalizing across images, COIN \textit{overfits} a neural network to a \textit{single} image. The quantized weights of this neural network (often referred to as a neural field \cite{xie2022neural}) are then transmitted as a compressed code for the image. As the decoder only needs to reconstruct a single image, the resulting network is significantly smaller than traditional neural decoders \cite{balle2018variational, minnen2018joint, cheng2020learned}, reducing the decoding complexity by orders of magnitude. However, while the decoding complexity of COIN is low, its RD performance is poor, and it is therefore not a viable alternative to other codecs.

More recently, \citet{ladune2023cool} introduced \cc which, in addition to learning a decoder per image like COIN, also learns an \textit{entropy} model per image. This led to significantly improved RD performance while maintaining low decoding complexity. 
A recent extension of \cc that we refer to as \cctwo
\cite{leguay2023low} exceeds the RD performance of the widely used BPG/HEVC codec \cite{sullivan2012overview, bellard2015bpg} while only requiring 2.3k MACs/pixel
at decoding time, an order of magnitude less than the most efficient neural codecs \cite{wang2023evc} (decoding complexity is measured in number of multiply-accumulate (MAC) operations, cf. \cref{app:sec:evaluation_details} for details). Despite these impressive results, the performance of \cc still falls short of the latest classical codecs such as VTM \cite{bross2021overview}. Further, \cc has not been applied to video, where low decoding cost is of greater importance as fast decoding is required to maintain a satisfactory frame rate for streaming.

In this paper, we introduce \method, a neural compression method that builds on \cc but substantially improves its RD performance while maintaining a low decoding complexity (see \cref{fig:bd-rate-vs-macs}). More specifically, we propose a series of simple and effective improvements to the optimization, quantization, and architecture of \cc. These changes result in a 22.2\% reduction in BD-rate \cite{bjontegaard2001calculation} compared to \cctwo while matching VTM on the CLIC2020 benchmark \cite{toderici2020workshop}. \textit{To the best of our knowledge, \method is the first neural compression method to achieve RD performance matching VTM on images while maintaining very low decoding complexity (less than 3k MACs/pixel)}. Further, \method is the state of the art among neural codecs obtained from a single image.

Going beyond COOL-CHIC, which is only applied to images, we also extend \method to videos, making several crucial methodological changes enabling the application of our method to this modality. On the UVG benchmark \cite{mercat2020uvg}, we demonstrate strong RD performance that matches VCT \cite{mentzer2022vct} while requiring 4.4k MACs/pixel, less than 0.1\% of VCT's decoding complexity. We believe this is a promising step towards efficient neural codecs trained on a single video.

\section{Background: \cc}
\label{sec:background}

Autoencoder based neural compression methods train an encoder network (also known as analysis transform) to compress an image $\vx$ into a quantized latent $\widehat\vz$, and a corresponding decoder network (also known as synthesis transform) to reconstruct $\vx$ from $\widehat\vz$. Typically, the latent $\widehat\vz$ is the only image-dependent component and is encoded into a bitstream using a shared entropy model $P$ \cite{yang2023introduction}.

In contrast, \cc \cite{ladune2023cool} and \cctwo \cite{leguay2023low} are methods for single image compression, in which all components are fit to each image separately.
In the following we provide further details on \cc. See \cref{fig:cc_model_overview} for an overview.

\paragraph{Overview.} At a high level, the \cc model consists of three components (cf. \cref{fig:cc_model_overview}):
(i) a set of latent grids at different spatial resolutions $\vz = (\vz^1, \dots, \vz^N)$, 
(ii) a synthesis transform \synthnet to decode these latents into an image, and 
(iii) an autoregressive entropy-coding network \entropynet that is used to convert the latents into a bitstream.
Because the networks do not need to be general, they can be very small, which allows for low decoding complexity.
Instead of an analysis transform, \cc uses optimization to jointly fit the latents, the synthesis transform and the entropy network per image. The gradient-based optimization acts on continuous values but is quantization-aware as we describe below; for the final encoding and decoding, the latent and the network parameters are both quantized.

\paragraph{Latent grids.} \cc structures the latent $\vz$ as a hierarchy of latent grids $(\vz^1, \dots, \vz^N)$ at multiple spatial resolutions to efficiently capture structure at different spatial frequencies. By default they are of shape $(h, w), (\tfrac{h}{2}, \tfrac{w}{2}), \dots, (\tfrac{h}{2^{N-1}}, \tfrac{w}{2^{N-1}})$, where $h$ and $w$ are the height and width of the image, respectively.

\paragraph{Synthesis.} The synthesis transform \synthnet approximately reconstructs the image $\vx$ from these latent grids. First, each latent grid $\vz^n$ is deterministically upsampled to the resolution of the image. 
Then, the synthesis network \synthnet uses the resulting concatenated tensor $\mathrm{Up}(\vz)$ of shape $(h, w, N)$ to predict the RGB values of the image, $\vx_\mathrm{rec} = \synthnet\left(\mathrm{Up}(\vz)\right)$ (see \cref{fig:cc_model_overview}{\small\sffamily{B}}). 
\cctwo uses learned upsampling and a small convolutional network to parameterize $\synthnet$.

\paragraph{Entropy coding.} For transmission, the latent grids and network parameters are quantized via rounding before being entropy-encoded into a bitstream. As this coding cost is dominated by the latent grids, an image-specific entropy model \entropynet is learned to losslessly compress them. %
\cc uses an integrated Laplace distribution for entropy coding, where the location and scale parameters $(\mu^n_{ij}, \sigma^n_{ij})$ of the distribution for each latent grid element $\vz^n_{ij}$ are autoregressively predicted by the entropy network from the local neighborhood of that grid element,
\begin{align}
    P_\psi(\vz^n) & = \textstyle\prod_{i, j} P\left(\vz^n_{ij}; \mu^n_{ij}, \sigma^n_{ij}\right)\\
    \mu^n_{ij}, \sigma^n_{ij} & = g_\psi\left(\mathrm{context}\left(\vz^n, (i, j)\right)\right).
\end{align}
Here, $\mathrm{context}(\vz^n, (i, j))$ extracts a small causally masked neighborhood ($5-7$ latent pixels wide) around a location $(i, j)$ from latent grid $\vz^n$ (c.f. \cref{fig:cc_model_overview}{\small\sffamily{A}}). Individual grids are modelled independently with the same network \entropynet, $P_\psi(\vz) = \textstyle\prod_n P_\psi(\vz^n)$.

The entropy and synthesis model are both small networks of depth $\leq 4$ and width $\leq 40$, and their parameters are quantized after training using different bin widths. The bin width with the best RD trade-off is chosen and added to the bitstream.
The quantized network parameters $\widehat\theta$ and $\widehat\psi$ are also entropy-coded using an integrated Laplace distribution that factorizes over entries with zero mean and scale determined by the empirical standard deviation:
\begin{align}
    P(\widehat\theta) & = \textstyle\prod_i P(\widehat\theta_i; \mu=0; \sigma=\tfrac{1}{\sqrt{2}}\mathrm{std}(\widehat\theta))
\end{align}
and similarly for $\widehat\psi$. Entropy coding for the latents and network parameters is performed using a range coder \cite{nigel1979rc}.

\paragraph{Quantization-aware gradient-based optimization.} 
The latent ($\vz$) and parameters ($\psi$, $\theta$) are fit to an image $\vx$ by jointly optimizing the following RD objective that trades off better reconstructions and more compressible latents with an RD-weight $\lambda$:
\begin{align}
    \mathcal{L}_{\theta, \psi}(\vz) = 
    \|\vx - \synthnet\left(\mathrm{Up}(\vz)\right)\|_2^2 - \lambda \textstyle\sum_n \log_2 P_\psi(\vz^n).
    \label{eq:rate_distortion_objective}
\end{align}

The optimization is made quantization-aware in several ways and proceeds in two stages (cf. \cref{tab:two_stages}): in the (longer) first stage, uniform noise $\vu$ is added to the continuous latents \vz; in the (shorter) second stage with very low learning rate, the latents $\vz$ are quantized and their gradients are approximated with the straight-through estimator, which is biased. Moreover, the rate term uses an integrated Laplace distribution.
\begin{table}[htb]
\centering
\begin{tabular}{ll}
     \toprule
     \small\sffamily \textcolor{gray}{Stage 1} & $\nabla_{\vz, \theta, \psi} \mathcal{L}_{\theta, \psi}(\vz + \mathbf{u}); \qquad \vu \sim\mathrm{Uniform}(0, 1)$ \\
     \small\sffamily \textcolor{gray}{Stage 2} & $\nabla_{\theta, \psi} \mathcal{L}_{\theta, \psi}(\lfloor\vz\rceil) \text{ and } \widetilde\nabla_{\vz} \mathcal{L}_{\theta, \psi}(\lfloor\vz\rceil)$\\
     \bottomrule
\end{tabular}
\vspace{-0.5em}
\caption{Two-stage optimization; $\widetilde\nabla_\vz$ is straight-through estimation.}
\label{tab:two_stages}
\end{table}
\section{\method: Improving COOL-CHIC}
\label{sec:method}

We first present a series of simple and effective improvements to COOL-CHIC, which we collectively refer to as \method (Cooler-ChiC), that lead to a significant increase in RD performance with similar decoding complexity. The overall model structure remains unchanged (cf. \cref{fig:cc_model_overview}) and most of our improvements fall into one of two categories: (1) improvements to the quantization-aware optimization, and (2) improvements to the model architecture.
Subsequently, we introduce the modifications necessary to apply \method to videos.
We confirm with extensive ablations in \cref{sec:results} and \cref{app:sec:additional_ablations} that each contribution is beneficial and that their improvements are cumulative.
See \cref{app:sec:method_details} for full details on all improvements.

\subsection{Optimization improvements}
\label{sec:method_opt}

We maintain the same two-stage optimization structure of COOL-CHIC (see \cref{tab:two_stages}) but make several improvements in both stages, most notably how quantization is approximated.

\begin{figure*}[ht]
    \centering
    \includestandalone[width=.975\textwidth, mode=image]{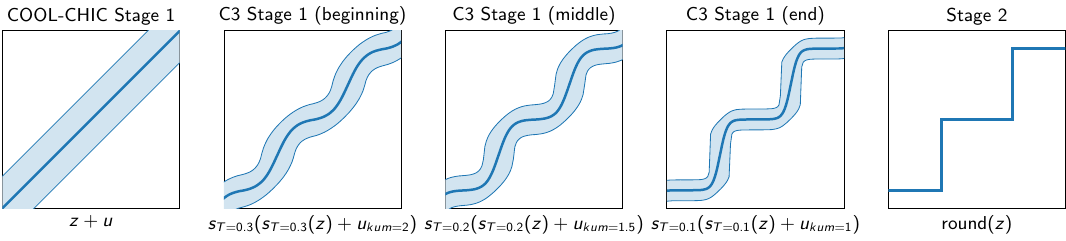}
    \caption{Approximating the $\mathrm{round}(\vz)$ function during Stage 1 of optimization. \cc adds uniform noise $\vu$, whereas \method uses soft-rounding $s_T$ with varying temperatures $T$ and Kumaraswamy noise of different strengths, $\vu_{\text{kum}}$. We plot the mean and $95\%$ interval.}
    \label{fig:rouning_noise}
\end{figure*}

\paragraph{Soft-rounding (stage 1).}
We apply a soft-rounding function before and after the addition of noise \cite{agustsson2020softrounddither}. Let $s_T$ be a smooth approximation of the rounding function whose smoothness is controlled by a temperature parameter $T$. For large $T$, $s_T$ approaches the identity while for small $T$, $s_T$ approaches the rounding function so that
\begin{align}
    \textstyle\lim_{T \rightarrow 0} s_T(s_T(\vz) + \mathbf{u}) = \lfloor \lfloor \vz \rceil + \mathbf{u} \rceil = \lfloor \vz \rceil.
\end{align}
By varying $T$ we can interpolate between rounding and the simple addition of uniform noise $\mathbf{u}$. Note that the soft-rounding does not create an information bottleneck as it is an invertible function. Therefore, adding noise is still necessary for reliable compression \cite{agustsson2020softrounddither}.

Small $T$ leads to a better approximation of rounding but increases the variance of gradients for $\vz$. Following previous work using soft-rounding, we therefore anneal the temperature over the course of the optimization. See \cref{fig:rouning_noise} for a visualization and \cref{app:sec:softrounding} for details.

\paragraph{Kumaraswamy noise (stage 1).}
The addition of uniform noise as an approximation to rounding has been motivated by pointing out that for sufficiently smooth distributions, the marginal distribution of the quantization error ($\vz - \lfloor \vz \rceil$) is approximately uniform~\cite{balle2016end}. The approximation further assumes that the quantization error and the input are uncorrelated. In practice, these assumptions may be violated, suggesting that other forms of noise are worth exploring. To that end, we replace uniform noise with samples from the Kumaraswamy distribution \cite{kumaraswamy1980} whose support is compact on $[0, 1]$. This distribution is very similar to the Beta distribution but has an analytic cumulative distribution function that allows for more efficient sampling. By controlling its shape parameters we can interpolate between a peaked (lower noise) distribution at beginning of stage 1 and a uniform distribution at the end. See \cref{fig:rouning_noise} for a visualization and \cref{app:sec:kumaraswamy} for details.

\paragraph{Cosine decay schedule (stage 1).}
We found that a simple cosine decay schedule for the learning rate of the Adam optimizer performed well during the first stage of optimization.

\paragraph{Smaller quantization step (stages 1 \& 2).}
COOL-CHIC quantizes the latents by rounding their values to the nearest whole integer; as a result the inputs to the synthesis and entropy networks can become large (exceeding values of $~50$), which can lead to instabilities or suboptimal optimization. We found that quantizing the latents in smaller steps than 1 (and correspondingly rescaling the soft-rounding in both stages) empirically improved optimization.

\paragraph{Soft-rounding for gradient (stage 2).}
We apply hard rounding/quantization to the latents \vz for the forward pass of stage 2 following COOL-CHIC. 
For the backward pass, \cctwo uses a straight-through gradient estimator and multiplies the gradient by a small $\epsilon$. This has the effect of replacing rounding by a linear function (cf. \cref{fig:rouning_noise}) and downscaling the learning rate of the latents. Instead we use soft-rounding to estimate the gradients (with a very low temperature) and start stage 2 with a low learning rate.

\paragraph{Adaptive learning rate (stage 2).}
We adaptively decrease the learning rate further when the RD-loss does not improve for a fixed number of steps.%

\subsection{Model improvements}
\label{sec:method_model}

We make a number of changes to the network architectures to increase their expressiveness, support the optimization, and allow for more adaptability depending on the bitrate.

\paragraph{Conditional entropy model.}
COOL-CHIC uses the same entropy network to independently model latent grids of starkly varying resolutions. We explored several options to increase the expressiveness of the entropy model: first, we optionally allow the context at a particular latent location to also include values from the previous grid, $P(\vz^n | \vz^{n-1})$, as this information may be helpful for prediction when different grids are correlated. Second, we optionally allow the network to be resolution-dependent by either using a separate network per latent grid or using FiLM \cite{perez2018film} layers to make the network resolution-dependent in a more parameter-efficient way.
\paragraph{ReLU $\rightarrow$ GELU.}
As we are constrained to use very small networks, we replace the simple ReLU activation function with a more expressive activation; empirically we found that GELU activations \cite{hendrycks2016gaussian} worked better. 

\paragraph{Shift log-scale of entropy model output.} Small changes in how quantities are parameterized can affect optimization considerably. For example, how the scale of the entropy distribution is computed from the raw network output strongly affects optimization dynamics, in particular at initialization; we found that shifting the predicted log-scale prior to exponentiation consistently improves performance.
With improved optimization we can also use larger initialization scales than COOL-CHIC to improve performance.

\paragraph{Adaptivity.} We optionally sweep over several architecture choices per image or video patch to find the best RD-trade-off on a per-instance basis. We refer to this as \emph{\method adaptive}. This setting includes an option to vary the relative latent resolutions; \eg, it may be beneficial not to use the highest resolution latent grid for low bitrates. Note that such adaptive settings are also common in traditional codecs \cite{itu1992jpeg, sullivan2012overview}.

\subsection{Video-specific methodology}
\label{sec:method_video}

\cc has been successfully applied to images but not videos. 
Here we describe our methodology for applying \method (and COOL-CHIC) to video, which we use on top of the improvements in \cref{sec:method_opt} and \cref{sec:method_model}.

\begin{figure}[tb]
    \centering
    \includestandalone[mode=image, width=\columnwidth]{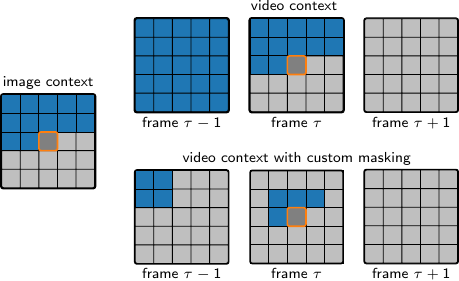}
    \caption{Visualization of entropy model's $\mathrm{context}$ for images and video (with and without custom masking).}
    \label{fig:video_context}
\end{figure}

\paragraph{2D $\rightarrow$ 3D.} Given that videos have an extra time dimension compared to images, a natural way to extend C3 to video is to convert 2D parameters and operations to their 3D counterparts. Namely, we use 3D latent grids $\vz^1, \vz^2, \ldots$ of shapes $(t, h, w), (\tfrac{t}{2}, \tfrac{h}{2}, \tfrac{w}{2}), \dots$, and the entropy model's context $\mathrm{context}(\vz^n, (\tau, i, j))$ is now a 3D causal neighborhood of the latent entry $\vz^n_{\tau ij}$ (cf. \cref{fig:video_context} $\mathrm{video \: context}$).

\paragraph{Using video patches.} 
Videos have orders of magnitude more pixels than images, and a full HD video does not fit into the memory of modern GPUs.
We therefore split the video into smaller video patches, and fit a separate \method model to each patch. We find that larger patches work best for lower bitrates and smaller patches work best for higher bitrates. Our patch sizes range from $(30, 180, 240)$ to $(75, 270, 320)$.

\paragraph{Wider context to capture fast motion.} For video patches with fast motion, the small context size that works well for images (5-7 latent pixels wide) can be smaller than the displacement of a particular keypoint in consecutive frames. This means that for a target latent pixel, the context in the previous latent frame does not contain the relevant information for the entropy model's prediction. Hence we use a wider spatial context (up to 65 latent pixels wide) to enhance predictions for videos with faster motion.

\paragraph{Custom masking.} Na\"{i}vely increasing the context width also increases the parameter count of the entropy model, which scales linearly with the context size. However, most of the context dimensions are irrelevant for prediction and can be masked out. We use a small causal mask centered at the target latent pixel for the current latent frame, and a small rectangular mask for the previous latent frame whose position is learned during encoding time (cf. \cref{fig:video_context} $\mathrm{video  \: context \: with \: custom \: masking}$). See \cref{app:sec:learned_custom_masking} for details of how the position of this mask is learned.

\section{Related work}
\label{sec:related-work}

\paragraph{Neural compression by overfitting to a single instance.} 
COIN \cite{dupont2021coin} introduced the idea of overfitting a neural network to a single image as a means for compression. This has since been improved with reduced encoding times through meta-learning \cite{strumpler2022implicit, dupont2022coin++, schwarz2022meta} and increased RD performance via better architectures \cite{catania2023nif} or more refined quantization \cite{damodaran2023rqat, gordon2023quantizing}. Further improvements to RD performance have been achieved by pruning networks \cite{lee2021meta, schwarz2022meta, ramirez2022_0} and incorporating traditional compressive autoencoders \cite{schwarz2023modality, pham2023autoencoding}. Recent approaches using Bayesian neural fields directly optimize RD losses, further improving performance \cite{guo2023compression, he2023recombiner}. Despite this progress, no approach yet matches the RD performance of traditional codecs such as VTM. 

For video, NeRV \cite{chen2021nerv} overfits neural fields to single videos, using a deep convolutional network to map time indices to frames. Various follow-ups have greatly improved compression performance \cite{chen2023hnerv, kwan2023hinerv, lee2023ffnerv, bai2023ps, li2022nerv, gomes2023video}, among which HiNeRV \cite{kwan2023hinerv} shows impressive RD performance that closely matches HEVC (HM-18.0, random access setting) on standard video benchmarks. While these models are typically smaller than autoencoder-based neural codecs, the model size (and hence decoding complexity) is directly correlated with the bitrate (each point on the RD curve corresponds to a different model size), making it challenging to design a low-complexity codec at high bitrates. Further, these models are typically unsuitable for video-streaming applications, as the entire bitstream needs to be transmitted before the first frame can be decoded \cite{van2021instance}. Note that \method does not suffer from this limitation -- the very small synthesis and entropy models can be transmitted first with little overhead, and then be used to decode the bitstream for the latents that can be synthesized into frames in a streaming fashion.

Given the generality of neural fields, codecs applicable to multiple modalities have been developed \cite{dupont2022coin++, schwarz2022meta, schwarz2023modality, girish2023shacira}. There also exist methods specialized to other modalities: climate data \cite{huang2022compressing}, 3D shapes \cite{davies2020effectiveness, lu2021compressive, isik2022lvac}, NeRF scenes \cite{takikawa2022variable, li2023compressing, girish2023shacira}, audio \cite{lanzendorfer2023siamese, guo2023compression} and medical images \cite{mancini2022lossy, dupont2022coin++, sheibanifard2023novel, gao2023sinco}.

\paragraph{Instance adaptive neural compression.} 
Several autoencoder-based approaches adapt the encoder to each instance through optimization but leave the decoder fixed \cite{campos2019content, guo2020variable, yang2020improving, lu2020content}. Such methods generally perform worse than approaches that optimize both the encoder and decoder w.r.t. an RD loss \cite{van2021overfitting, van2021instance, lv2023dynamic, mikami2021efficient}. 
In particular, \citet{van2021instance} introduce Insta-SSF, an instance adaptive version of the scale-space flow (SSF) model \cite{agustsson2020scale} (a popular autoencoder model for neural video compression). For a fixed RD performance, the decoder of Insta-SSF is much smaller and has lower complexity than the shared decoder of SSF. Note that \method and COOL-CHIC follow the same principle for low complexity decoding. 
However, there are key differences between \method/\cc and the aforementioned instance adaptive methods: 1. we train from scratch rather than learning an initialization from a dataset; 2. we use a neural field model (without encoder) instead of an autoencoder, and show an order of magnitude lower decoding complexity; 3. for videos, there is no explicit motion compensation based on flows in our model.

\begin{figure}[t]
  \centering
   \includestandalone[mode=image, width=\columnwidth]{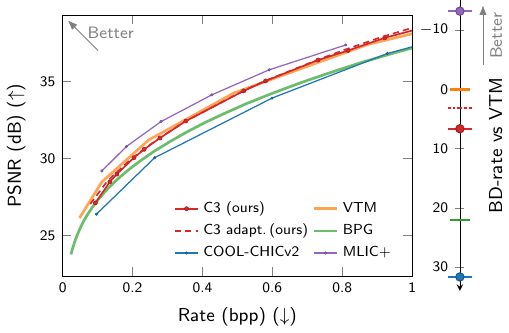}
   \caption{Rate-distortion curve and BD-rate on Kodak.}
   \label{fig:rd-curve-kodak}
\end{figure}

\paragraph{Low complexity neural codecs.}
While the problem of high decoding complexity in neural compression is well established \cite{yang2023introduction}, most works to mitigate it are relatively recent. Early methods reduced complexity at little cost in RD performance by pruning network weights \cite{johnston2019computationally}. More recently, \citet{he2021checkerboard, he2022elic} replace traditional autoregressive entropy models with checkerboard-based designs that allow for more efficient and parallelizable entropy coding. Further, \citet{yang2023computationally} use shallow decoders to reduce decoding complexity and offset the resulting decrease in RD performance with iterative encoding. EVC \cite{wang2023evc} achieves RD performance surpassing VTM on images with decoding at 30FPS on a GPU, by carefully choosing architectures and using sparsity-based mask decay. Despite these impressive results, the decoding complexity required for these models is still an order of magnitude higher than \method. 

For video, some prior works \cite{le2022mobilecodec, van2023mobilenvc} focus on providing efficient neural components and entropy coding that run on mobile devices. Due to these constraints, their RD performance is not yet competitive with most neural video codecs and their decoding complexity is an order of magnitude higher than \method. 
ELF-VC \cite{rippel2021elf}, based on autoencoders and flows, provides gains in efficiency by encoder/decoder asymmetry and an efficient convolutional architecture. However they do not report decoding complexity and are outperformed by VCT \cite{mentzer2022vct} in terms of RD. AlphaVC \cite{shi2022alphavc} introduces a technique to skip latent dimensions for entropy coding, improving efficiency in flow-based autoencoder models and surpassing VTM (low-delay) in terms of RD performance, albeit with a high decoding complexity of $1$M MACs/pixel.

\section{Results}
\label{sec:results}

\subsection{Image compression}

We evaluate our model on the Kodak \cite{kodakdataset} and CLIC2020 \cite{toderici2020workshop} benchmarks. Kodak contains $24$ images at a resolution of $512 \times 768$. For CLIC2020, we use the professional validation dataset split containing $41$ images at various resolutions from $439 \times 720$ to $1370 \times 2048$, following \cc \cite{ladune2023cool, leguay2023low}. We compare \method against a series of baselines, including classical codecs (BPG \cite{bellard2015bpg}, VTM \cite{bross2021overview}), autoencoder based neural codecs (BMS \cite{balle2018variational}, a standard neural codec; CST \cite{cheng2020learned}, a strong neural codec; EVC \cite{wang2023evc}, a codec optimized for RD performance and low decoding complexity; MLIC+ \cite{jiang2022multi}, the state of the art in terms of RD performance) and COOL-CHICv2 \cite{leguay2023low}. We measure PSNR on RGB and quantify differences in RD performance with the widely used BD-rate metric. See \cref{app:sec:method_details} for full experimental details and \cref{app:sec:evaluation_details} for full evaluation details.

\begin{figure}[t]
  \centering
   \includestandalone[mode=image, width=\columnwidth]{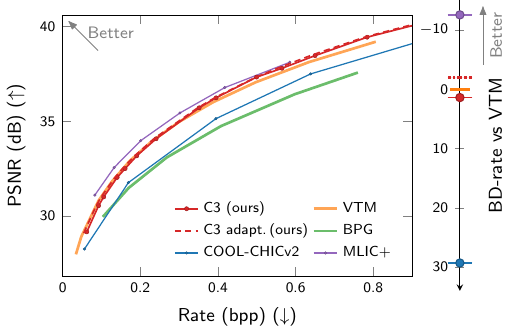}
   \caption{Rate-distortion curve and BD-rate on CLIC2020.} 
   \label{fig:rd-curve-clic}
\end{figure}

\paragraph{Rate-distortion and decoding complexity.} On CLIC2020, \method (with a single setting for its architecture and hyperparameters) significantly outperforms \cctwo across all bitrates ($-22.2\%$ BD-rate) and nearly matches VTM ($+1.4\%$ BD-rate), cf. \cref{fig:rd-curve-clic}. When adapting the model per image, \method even outperforms VTM ($-2.0\%$ BD-rate). \emph{To the best of our knowledge, this is the first time a neural codec has been able to match VTM while having very low decoding complexity (below 3k MACs/pixel).} While \method does not yet match the RD performance of state of the art neural codecs such as MLIC+, it uses two orders of magnitude fewer operations at decoding time, making it substantially cheaper.
Results are also strong on Kodak (see \cref{fig:rd-curve-kodak}), although, as is the case for COOL-CHIC, we perform slighly worse on this dataset relative to VTM. 
In \cref{fig:bd-rate-vs-macs} we compare the decoding complexity (measured in MACs/pixel) and the achieved BD-rate for \method and other neural baselines. \method has a similar complexity to \cc but much better BD-rate and codecs achieving similar BD-rate to ours require at least an order of magnitude more MACs (even ones optimized for low decoding complexity such as EVC). %
See \cref{app:sec:full_rd_curves} for comparisons with additional baselines (including other autoencoder based codecs and overfitted codecs) in terms of RD performance and decoding complexity.

\paragraph{Decoding time.}
A concern with using autoregressive models is that runtimes may be prohibitive despite low computational complexity \cite{minnen2023advancing}. To address this, we time the decoding process, which includes a full iterative roll-out of the autoregressive entropy model (and the upsampling and application of the synthesis network). On CPU (Intel Xeon Platinum, Skylake, 2GHz) these together take ${}<100$\,ms ($\sim55$\,ms and $\sim30$\,ms, respectively) for an image of size $768 \times 512$. This does not account for the cost of range-decoding the bitstream (which is also a component of every classical codec). We emphasize that these numbers are based on unoptimized research code and can likely be improved substantially.

\paragraph{Encoding time.}
\method faces the same limitations as COOL-CHIC, in that it has very long encoding times. Here we report encoding times on an NVIDIA V100 GPU. The largest CLIC image at $1370 \times 2048$ resolution takes 48s per 1000 iterations of optimization (\ie, excluding range-encoding) with the slowest setting (largest architecture), and 22s per 1000 iterations with the fastest setting (smallest architecture). While we train for a maximum of 110k iterations, we show in \cref{app:sec:ablation:encoding_iterations} that we can approach similar RD performance with much fewer iterations. As we run unoptimized research code, we believe the runtime can be greatly improved.

\paragraph{Ablations.} In \cref{tab:ablations}, we ablate our methodological contributions on Kodak by starting with our best performing model and sequentially removing each of our improvements, one after another.
We show the resulting BD-rate with respect to the top row, demonstrating that our contributions stack to yield significant improvements in RD performance. In \cref{tab:ablation_knockout}, we show BD-rate with respect to \method when disabling individual features. We find that soft-rounding, Kumaraswamy noise and using GELU activations are responsible for the majority of the improvement.
For the corresponding ablations on CLIC2020, please refer to \cref{app:sec:clic_ablations}.

\begin{table}[ht]
    \centering
    \includestandalone[mode=image, width=\columnwidth]{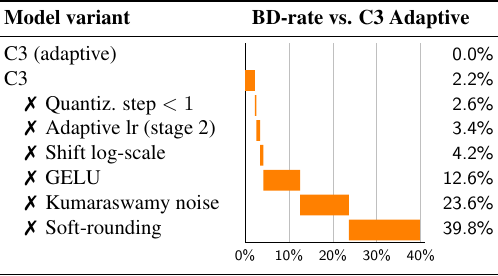}
  \caption{Kodak ablation \textit{sequentially} removing one improvement after another. Note higher BD-rate means worse RD performance.}
  \label{tab:ablations}
\end{table}

\begin{table}[ht]
    \centering
    \begin{tabular}{lC}
    \toprule
    \textbf{Removed feature} & \textbf{BD-rate vs. \method} \\
    \midrule
    Soft-rounding & 22.18\% \\ %
    Kumaraswamy noise & 3.90\%\\ %
    GELU & 3.27\% \\ %
    Shifted log-scale & 0.87\% \\ %
    Adaptive lr (stage 2) & 0.68\% \\ %
    Quantization step${}< 1$ & 0.40\% \\  %
    \bottomrule
    \end{tabular}
    \caption{Kodak ablation knocking out individual features from \method (fixed hyperparameters across all images). Note higher BD-rate means worse RD performance.}
    \label{tab:ablation_knockout}
\end{table}

\begin{figure}[t]
  \centering
   \includestandalone[mode=image, width=0.975\columnwidth]{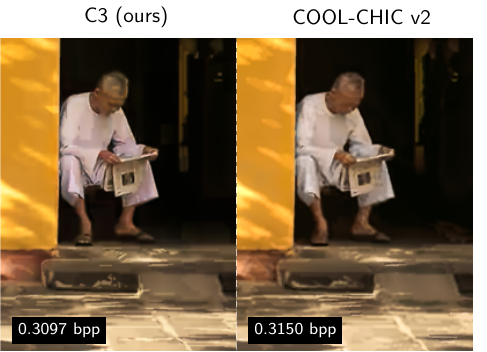}
   \caption{Qualitative comparison of compression artifacts for \method and COOL-CHICv2 at around $0.31$\,bpp with a PSNR of 30.28dB and 28.98dB, respectively. See \cref{app:sec:additional_visualizations} for the full image.}
   \label{fig:qualitative}
   \vspace{-3mm}
\end{figure}

\paragraph{Qualitative comparisons} In \cref{fig:qualitative}, we compare reconstructions from \method and COOL-CHICv2 on an image from CLIC2020, showing that \method has fewer artifacts. See \cref{app:sec:additional_visualizations} for a more thorough comparison. 

\subsection{Video compression}
\label{sec:video_compression}

We evaluate \method on the UVG-1k dataset \cite{mercat2020uvg} containing 7 videos at HD resolution ($1080 \times 1920$) with a total of 3900 frames. We evaluate PSNR on RGB, and compare against a series of baselines, including classical codecs (HEVC medium, no B-frames \cite{sullivan2012overview}), neural codecs based on overfitting (HiNeRV \cite{kwan2023hinerv}, FFNeRV \cite{lee2023ffnerv}) and autoencoder based neural codecs (DCVC \cite{li2021deep}, VCT \cite{mentzer2022vct}, Insta-SSF \cite{van2021overfitting}, MIMT \cite{xiang2022mimt}), among which MIMT reports state of the art RD performance on the UVG-1k dataset. Note that extensions of DCVC \cite{sheng2022temporal, li2022hybrid, li2023neural} also show strong RD performance but report results on a subset of UVG frames, hence we do not compare against them. See \cref{app:sec:method_details} for full experimental details and \cref{app:sec:evaluation_details} for full evaluation details.

\begin{figure}[t]
  \centering
   \includestandalone[mode=image, width=\columnwidth]{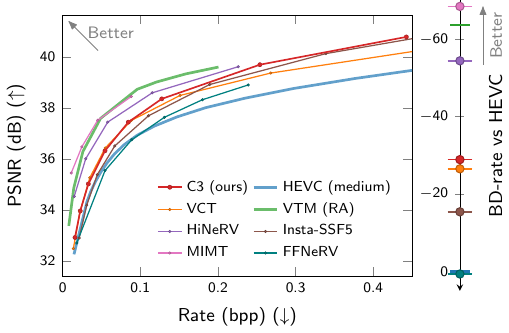}
   \caption{Rate-distortion curve and BD-rate on UVG.}
   \label{fig:rd-curve-uvg}
\end{figure}

\begin{figure}[t]
  \centering
  \includestandalone[mode=image, width=\columnwidth]{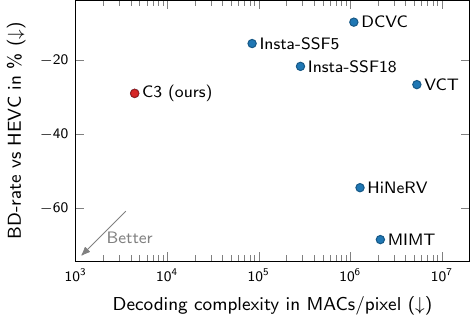}
  \caption{BD-rate vs decoding complexity relative to HEVC (medium). For methods with varying MACs for different bitrates (\eg, \method and HiNeRV), we report the largest MACs/pixel.}
   \label{fig:bd-rate-vs-macs-uvg}
   \vspace{-3mm}
\end{figure}

\paragraph{Rate-distortion and decoding complexity} In \cref{fig:rd-curve-uvg}, we show the RD performance of \method compared to other baselines, with more baselines shown in \cref{app:sec:full_rd_curves}. In \cref{fig:bd-rate-vs-macs-uvg}, we show the MACs/pixel count of each method vs the BD-rate using HEVC (medium, no B-frames) \cite{sullivan2012overview} as anchor. In terms of RD performance, we are on par with VCT \cite{mentzer2022vct}, a competitive neural baseline, while requiring 4.4k MACs/pixel, which is less than 0.1\% of VCT's MACs/pixel. Among the baselines that overfit to a single video instance (NeRV and its followups) we are second best in terms of RD, widely outperforming FFNeRV \cite{lee2023ffnerv}, the previous runner up. Although \method is behind stronger neural baselines such as HiNeRV and MIMT in terms of RD performance, our decoding complexity is orders of magnitude lower. Note that NeRV-based methods have different model sizes (and hence different MACs/pixel) for each point on the RD curve. For example, the 5 points on the RD curve for HiNeRV correspond to MACs/pixel values between 87k-1.2M \cite{kwan2023hinerv}. In \cref{app:sec:video_ablations} we show ablation studies showing the effectiveness of our video-specific methodology.

\paragraph{Encoding times.}
We also report encoding times for video patches on an NVIDIA V100 GPU. The slowest setting on the largest video patch of size $75 \times 270 \times 320$ resolution takes 457s per 1000 iterations of optimization, whereas the fastest setting on the smallest video patch of size $30 \times 180 \times 240$ takes 29s per 1000 iterations. We train for a maximum of 110k iterations but show in \cref{app:sec:ablation:encoding_iterations} that we can approach similar RD performance with much fewer iterations.
\section{Conclusion, limitations and future work}
\label{sec:discussion}

We propose \method, the first low complexity neural codec on single images that is competitive with VTM while requiring an order of magnitude fewer MACs for decoding than state of the art neural codecs.
We then extend \method to the video setting, where we match the RD performance of VCT with less than 0.1\% of their decoding complexity. 
Our contributions are a step towards solving one of the major open problems in neural compression --- achieving high compression performance with low decoding complexity --- and ultimately towards making neural codecs a practical reality.

\paragraph{Limitations.} In this paper, we focused on maximizing RD performance while minimizing decoding complexity. As a result, the encoding of \method is slow, making it impractical for use cases requiring real time encoding. Yet, there are several use cases for which paying a significant encoding cost upfront can be justified if RD performance and decoding time are improved. For example, a popular video on a streaming service is encoded once but decoded millions of times \cite{netflix-optim}.
Further, the autoregressive entropy model used during decoding is inherently sequential in nature, posing challenges for efficient use of hardware designed for parallel computing.
However, as shown in \cref{sec:results}, even with unoptimized research code, an image can be decoded relatively quickly on CPU due to the very small network sizes. Moreover, further optimizations and specialized implementations such as wavefront decoding \cite{clare2011wavefront} can likely speed up decoding times significantly. Nevertheless, it would be interesting to explore alternative probabilistic models that can be efficiently evaluated on relevant hardware.

\paragraph{Future work.} There are several promising avenues for future work. Firstly, it would be interesting to accelerate encoding via better initializations or meta-learning \cite{strumpler2022implicit, dupont2022coin++, schwarz2022meta}. Secondly, improving decoding speed through the use of different probabilistic models or decoding schemes is an important direction. Further, while we took an extreme view of using only a single image or video to train our models, it is likely that some level of sharing across images or videos could be beneficial. For example, sharing parts of the entropy or synthesis model may improve RD performance. %

\section*{Acknowledgements}
We would like to thank: Wei Jiang for providing RD and MACs/pixels numbers for several baselines both on the Kodak and CLIC2020 datasets; Fabian Mentzer and Ho Man Kwan for providing video RD numbers for various baselines; Eirikur Agustsson for helping to run VTM (RA); COOL-CHIC authors for open sourcing their code; Yee Whye Teh, Nick Johnston, Fabian Mentzer and Eirikur Agustsson for helpful feedback.

\section*{Author Contributions}
ED conceived the project and wrote the initial codebase with the help of HK and MB. 
HK, MB, ED, LT developed and refined the project vision with the help of JRS. 
MB, LT, HK, ED  implemented and refined the general methodology. 
HK designed and implemented the video-specific methodology and evaluation with help from ED and MB. 
MB, HK, LT worked on scaling, evaluating and improving the efficiency of experiments. 
MB, ED, HK, LT wrote the paper. 

These authors contributed equally: HK, MB, ED.

{
    \small
    \bibliographystyle{ieeenat_fullname}
    \bibliography{main}
}

\clearpage
\appendix
\onecolumn
\maketitlesupplementary

\section{Method and experimental details}
\label{app:sec:method_details}

\subsection{Model architecture}

Here we provide full details of all model components of \method, cf. \cref{fig:cc_model_overview} in the main paper for a visualization of the model. In \cref{app:sec:hyperparameters} we provide all hyperparameter settings for our experiments.

\subsubsection{Multi-resolution latent grids}
We follow \cc and structure the latent $\vz$ in a hierarchy of $N$ latent grids, $\vz^1, \dots, \vz^N$, at multiple resolutions. Each latent grid $\vz^n$ has a single channel and is of shape $(h_n, w_n)$ for images and $(t_n, h_n, w_n)$ for videos. By default, these sizes are related to the image shape $(h, w)$ or video shape $(t, h, w)$ through
\begin{align}
    (t_n, h_n, w_n) & = (\tfrac{t}{2^{n-1}}, \tfrac{h}{2^{n-1}}, \tfrac{w}{2^{n-1}}), \qquad n=1, \dots, N,
\end{align}
that is, the latent grid $\vz^n$ is a factor $2$ smaller in each dimension than the previous grid $\vz^{n-1}$.
All latent grids are initialized to zero at the start of optimization.

\subsubsection{Upsampling the latent grids to the resolution of the input}
Each latent grid is deterministically upsampled to the input resolution $(\{t\}, h, w)$ before all grids are concatenated together and passed as input to the synthesis network. \cc uses simple bicubic upsampling for this \cite{ladune2023cool}. \cctwo instead uses learned upsampling that is implemented as a strided convolution (allowing for upsampling by a factor of $2$ only) that is initialized to bicubic upsampling \cite{leguay2023low}.

\begin{landscape}
    \begin{figure}
        \centering
        \includestandalone[mode=image]{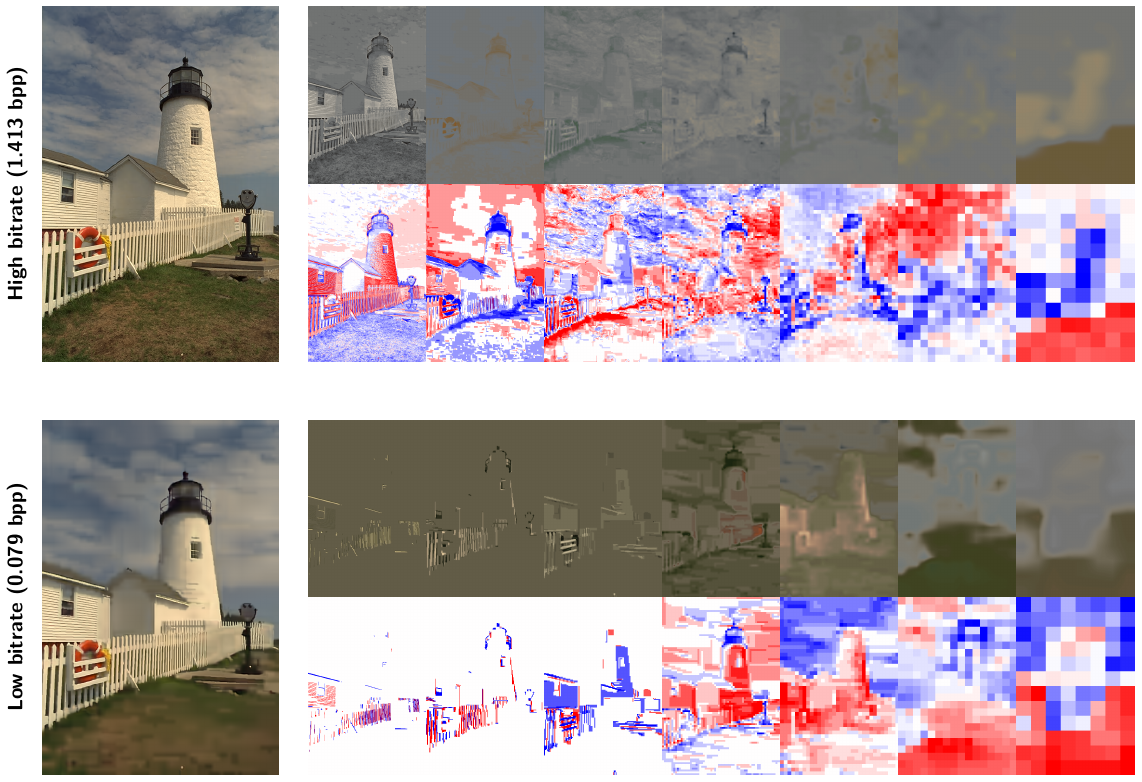}
        \caption{\textit{Top:} Reconstruction and visualization of C3's latents for kodim19 at a high bit-rate (1.413 bpp). The first row shows reconstructions when all but one out of 7 sets of latents is set to zero. For example, the highest resolution latent grid appears to encode luminance information. The second row visualizes the raw latents, upscaled to match the resolution of the output. \textit{Bottom:} As above but at a lower rate (0.079 bpp).}
        \label{app:fig:latents}
    \end{figure}
\end{landscape}

We experimented with different forms of upsampling (both learned and fixed) in our setup but found that for almost all bitrates, using simple bilinear upsampling led to the best results. More complex upsampling methods such as bicubic or Lanczos upsampling only led to better results for very low bitrates. We explain this observation as follows. For high bitrates, fine details are modeled by the highest resolution latent grid, which already matches the resolution of the input, see, \eg, the top row in \cref{app:fig:latents}. Therefore bilinear upsampling of the lower resolution latents is sufficient. For low bitrates, only very few details are modeled by the highest resolution latent grid, see bottom row in \cref{app:fig:latents}. The model therefore relies much more heavily on information upsampled from the lower resolution latent grids to explain fine details; in this case, more complex upsampling methods are beneficial.
Because the differences even for lowest bitrates were very small, we opted to exclusively use bilinear interpolation as it also has a lower decoding complexity.
For videos we use trilinear interpolation.

\subsubsection{Image synthesis with the synthesis network}

The upsampled latents are stacked into a single tensor of shape $(\{t\}, h, w, N)$ and are then used as input for the synthesis network \synthnet, which directly predicts the raw RGB intensity values (output values are clipped to lie in the correct range). 

To parameterize \synthnet, \cc uses a simple MLP that is applied separately to each of the $(\{t\}, h, w)$ pixel locations. This operation can be equivalently implemented as a sequence of $1 \times 1$ convolutions. In addition, \cctwo optionally adds several $3\times 3$ residual convolutions. For the residual convolutions the input and output channel dimensionality is set to $3$ to keep the decoding complexity low. \method follows the same architecture layout but uses the more expressive GELU \cite{hendrycks2016gaussian} activation function instead of ReLUs. We also opt to use narrower and deeper networks with a similar overall decoding complexity. For \method, the $1 \times 1$ convolutions are initialized with the standard He initialization while the residual convolutions are initialized to zero.

\subsubsection{Entropy model for the latent grids}
\label{app:sec:entropy_model_details}
\begin{figure}[tb]
    \centering
    \includestandalone[mode=image]{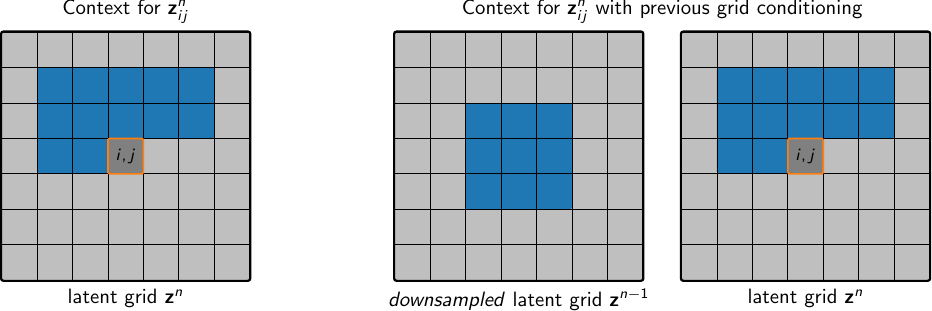}
    \caption{Illustration of the context used by the entropy model to predict the distribution parameters of a latent grid location $\vz^n_{ij}$. \textit{Left:} Without previous grid conditioning the context only comes from a (causal) neighborhood in the same latent grid. \textit{Right:} With previous grid conditioning a small neighborhood from the bilinearly downsampled latent grid $\vz^{n-1}$ is used in addition to the context from the current latent grid $\vz^n$.}
    \label{app:fig:prev_grid_conditioning}
\end{figure}

The entropy model is used to losslessly compress the (quantized) latent grids into a bitstream. Each latent grid location, $\vz^n_{ij}$ for images and $\vz^n_{\tau ij}$ for videos, respectively, is entropy en-/decoded using a quantized Laplace distribution with mean parameter $\mu^n_{ij}$ ($\mu^n_{\tau ij}$ for video) and scale parameter $\sigma^n_{ij}$ ($\sigma^n_{\tau ij}$ for video).
Both parameters are autoregressively predicted from the context of the latent grid entry using the entropy network \entropynet:
\begin{align}
    \mu^n_{ij}, \sigma^n_{ij} & = g_\psi\left(\text{context}\left(\vz^n, (i, j)\right)\right). && \text{for images} \\
    \mu^n_{\tau ij}, \sigma^n_{\tau ij} & = g_\psi\left(\text{context}\left(\vz^n, (\tau, i, j)\right)\right). && \text{for videos}
\end{align}
Because the autoregressive prediction also occurs at decoding time, the context has to be causally masked/extracted; following \cc, we use a (causally masked) neighborhood of the latent entry as its context (cf. \cref{app:fig:prev_grid_conditioning} \textit{(left)}).
\entropynet is a fully connected network that maps the context to the Laplace distribution parameters. 
During optimization and encoding, when all latent entries are available, we use $1\times 1$ convolutions ($1\times 1 \times 1$ for video) with the corresponding number of channels to replace the MLP and use a masked $k_h \times k_w$ convolution ($k_t \times k_h \times k_w$ for video) to replace the extraction of context of size $k$ and the first layer of the MLP, cf. \cref{app:fig:prev_grid_conditioning} \textit{(left)}.

The scale parameter $\sigma$ of the Laplace distribution is constrained to be positive while the output of the entropy network \entropynet in unconstrained and can be positive and negative. We therefore pass the raw prediction of the network through an exponential function; in other words, the network predicts the log-scale instead of the scale as is usually done in practice when parameterizing positive values. We found that shifting the predicted raw log-scale value by a constant can improve optimization dynamics as it determines the behavior of the model close to initialization.

So far the context for an entry $\vz^n_{ij}$ only takes into account the local neighborhood in the same grid $\vz^n$. This means that the entropy model factorizes across grids:
\begin{align}
    P_\psi(\vz) = \prod_n P_\psi(\vz^n).
\end{align}
Alternatively, we also consider the option of extending this context to include local grid entries in a neighboring grid $\vz^{n-1}$:
\begin{align}
    P_\psi(\vz) = P_\psi(\vz^1) \prod_{n>1} P_\psi(\vz^n|\vz^{n-1}).
\end{align}
Note that because of the autoregressive structure, we can only depend on the neighboring grid in one direction. In that case, we downsample the previous latent grid $\vz^{n-1}$ to the same resolution as the current grid, $\vz^n$, and extract the full neighborhood around the location of interest from it; cf. \cref{app:fig:prev_grid_conditioning} \textit{(right)} for an illustration. The size of the neighborhood extracted from the previous grid can vary from the size of the neighborhood in the current grid. We found that in practice, small neighborhood sizes for the previous grid (\eg $3\times 3$) were sufficient.

We also explored the option of using separate entropy parameters for different grids
\begin{align}
    P_\psi(\vz) = \prod_n P_{\psi^n}(\vz^n),
\end{align}
and found that while this did not help for images, it gave better RD performance on videos, especially when using different masking patterns for different grids (cf. \cref{app:sec:learned_custom_masking}).

\subsection{Quantization-aware optimization of the rate-distortion objective}

Here, we provide further details about the RD objective in \cref{eq:rate_distortion_objective} that is used to fit the latent grids $\vz$ as well as the synthesis network \synthnet and the entropy network \entropynet to a particular image $\vx$ using gradient-based optimization. We reproduce the objective here for easier reference:
\begin{align}
    \mathcal{L}_{\theta, \psi}(\vz) = 
    \|\vx - \synthnet\left(\mathrm{Upsample}(\vz)\right)\|_2^2 - \lambda \log_2 P_\psi(\vz).
    \tag{\ref{eq:rate_distortion_objective}}
\end{align}
The objective trades off reconstruction of the image (first term) and compression of the latent \vz (second term) with an RD-weight $\lambda$. By varying $\lambda$ we trace out different points in the RD-plane that we plot as RD-curves. High values of $\lambda$ lead to low bitrates and vice versa.
Note that the objective does not take into account the quantization of the model parameters $\theta$ and $\psi$.

\

At evaluation time, the distortion (measured in PSNR) and the rate are computed from quantized versions of the variables, $\widehat\vz, \widehat\theta$, and $\widehat\psi$, that have been entropy-decoded from the bitstream, see \cref{fig:cc_model_overview} for an illustration of the decoding. 

During optimization we evaluate the RD-loss in \cref{eq:rate_distortion_objective} using continuous variables $\vz, \theta, \psi$. Naive minimization of the objective w.r.t. the continuous variables would give rise to a solution that does not work well when the variables are quantized. Instead, we have to take the subsequent quantization into account during optimization. In practice, quantization of the latent grid values $\vz$ is most relevant, such that the optimization is only made aware of the quantization of the latent grids and not of the quantization of the network parameters themselves. We explain how these network parameters are quantized and entropy en-/decoded in \cref{app:sec:method:quantize_network_params}.

\subsubsection{Quantized Laplace distribution for continuous variables: Integrated Laplace distribution}

As explained above, the rate is modeled by (the log density of) a quantized Laplace distribution $P_\psi$ whose distribution parameters are predicted by the entropy model \entropynet. 

During optimization we use continuous values and, following \citet{balle2018variational} and \citet{ladune2023cool}, replace the quantized Laplace distribution with an integrated Laplace distribution that integrates the probability mass over the rounding/quantization interval. When evaluated on quantized values, the two distributions are identical:
\begin{align}
    P_\psi(\vz^n_{ij}) & = \int_{\vz^n_{ij}-0.5}^{\vz^n_{ij}+0.5} \mathrm{Laplace}(z; \mu^n_{ij}, \sigma^n_{ij}) \, \mathrm{d}z,
\end{align}
where the location parameter $\mu^n_{ij}$ and the scale parameter $\sigma^n_{ij}$ of the Laplace distribution are autoregressively predicted by the entropy network \entropynet.

\subsubsection{Two stages of optimization}

Following \cc and \cctwo, we split the optimization into two stages that differ in how the quantization of the latents is approximated. As discussed in the main paper, we make improvements to both stages. Here, we explain the two stages used by \method in details; we also highlight the main differences to prior work where appropriate.
\begin{description}
\item[Stage 1: Soft-rounding $\vz$ and adding noise to it.] In the first stage, the continuous values for the latent grids $\vz$ are passed through an invertible soft-rounding function and additionally perturbed with additive noise as we describe in \cref{app:sec:softrounding,app:sec:kumaraswamy}, respectively. Because the soft-rounding function is invertible and differentiable everywhere, we can compute its gradient with backpropagation, and the additive noise can be ignored for the gradient computation as it does not depend on any of the parameters (the soft-rounding function is implemented in a reparameterized form):
\begin{align}
    \text{forward}: \qquad & \mathcal{L}_{\theta, \psi}(\mathrm{softround}_T(\vz, \vu)) && \vu\sim p_\text{noise}(\vu) \label{app:eq:method:stage1_objective}\\
    \text{backward for } \theta, \psi: \qquad & \nabla_{\theta, \psi}\,\mathcal{L}_{\theta, \psi}(\mathrm{softround}_T(\vz, \vu)) \label{app:eq:method:stage1_net_grads}\\
    \text{backward for } \vz: \qquad & \nabla_{\vz} \, \mathcal{L}_{\theta, \psi}(\mathrm{softround}_T(\vz, \vu)) \label{app:eq:method:stage1_z_grads}
\end{align}
Gradient variance is a concern in this stage of training, especially since we use larger learning rates. Because of this, we cannot use a temperature $T$ in the soft-rounding that is too low (cf. \cref{app:sec:softrounding} for details), and we also found it beneficial to use more concentrated noise distributions than uniform noise early in training (cf. \cref{app:sec:kumaraswamy} for details). 

\cc and \cctwo do not use soft-rounding and instead directly add uniform noise, $p_\text{noise}(\vu) = \mathrm{Uniform}(\vu)$.
\item[Stage 2: Hard-rounding $\vz$.] In the second stage, the continuous values for the latent grids $\vz$ are (hard-)rounded; \ie, they are replaced by their quantized values $\widehat\vz = \lfloor \vz \rceil$. Quantizing the latents increases the variance of the gradients w.r.t. the network parameters $\theta$ and $\psi$ and necessitates lower learning rates as we discuss in \cref{app:sec:schedules}. To estimate gradients w.r.t. the latent $\vz$, we have to backpropagate through the discrete rounding; as this is not possible, we replace the hard-rounding with soft-rounding using a very low temperature $T=10^{-4}$ for this case. This estimator approximates the hard-rounding well but is invertible; however, it is still biased. Note that we do not add any noise when using this soft-rounding estimator.
\begin{align}
    \text{forward}: \qquad & \mathcal{L}_{\theta, \psi}(\lfloor \vz \rceil)\\
    \text{backward for } \theta, \psi: \qquad & \nabla_{\theta, \psi}\,\mathcal{L}_{\theta, \psi}(\lfloor \vz \rceil) \label{app:eq:method:stage2_net_grads} \\
    \text{backward for } \vz: \qquad & \nabla_{\vz} \, \mathcal{L}_{\theta, \psi}(\mathrm{softround}_{T_{\rightarrow 0}}(\vz)) \label{app:eq:method:stage2_z_grads}
\end{align}

Because of our improvements to stage 1, the loss as well as the corresponding rate and distortion values do not change by much in the second stage of optimization. Overall, stage 2 seems to be less important for \method than for \cctwo, though it still leads to small improvements. See \cref{app:sec:ablation:stage_2} for an ablation.

\cc also uses the quantized latent $\widehat\vz$ in the second stage but uses simple (linear) straight-through estimation to estimate gradients w.r.t. $\vz$ \cite{ladune2023cool}; the linear function is a cruder approximation of (hard) rounding than the soft-rounding function, such that the bias of this estimator is larger than for \method.
\cctwo also uses linear straight-through estimation but downscales the gradient by a factor $\epsilon \ll 1$ \cite{leguay2023low}. This results in the same biased straight-through estimator but effectively changes the learning rate of the latents to be smaller.
\end{description}

\subsubsection{Learning rate decay}
\label{app:sec:schedules}

\begin{figure}[htb]
    \centering
    \includestandalone[mode=image]{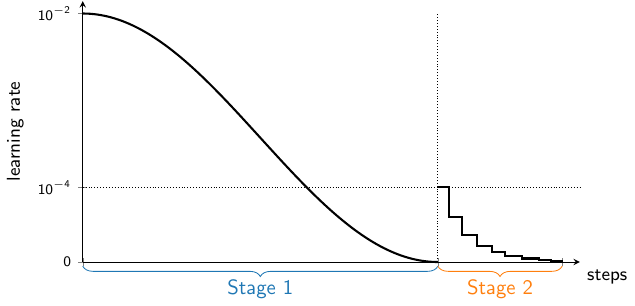}
    \caption{Schematic of the learning rates used for the two stages of optimization. Axes are not to scale and values are only indicative. In stage 1, the learning rate is decayed with a cosine schedule from an initial value to $0$ at the end of stage 1. In stage 2 we adaptively decay the learning rate by a constant factor whenever the evaluation loss does not improve for a certain number of steps.}
    \label{app:fig:lr_schematic}
\end{figure}

As for most optimization based algorithms the learning rate is one of the most important hyperparameters in \method. We use two simple strategies to choose the learning rate for stage 1 and stage 2, respectively, cf. \cref{app:fig:lr_schematic} for a schematic.

\begin{description}
    \item[Learning rate in Stage 1.] We use a cosine decay schedule that starts at a higher value and is decayed to $0$ throughout stage 1, which also makes up most of the optimization steps. The initial learning rate value is chosen empirically.
    \item[Learning rate in Stage 2.] Due to the variance of the gradients and the bias of the estimator, the second stage of optimization depends even more strongly on the learning rate. We found that starting with a high enough learning rate was important to make progress, but that an aggressive decay of the learning rate may be necessary as otherwise the loss can quickly get worse. Instead of using a fixed schedule, we therefore opted for the following automatic and adaptive mechanism: Starting from a fixed sufficiently high learning rate ($10^{-4}$ in our experiments), we track the loss and decay the learning rate by a fixed factor if the loss does not improve for a certain number of steps. Upon decaying the learning rate we also reset the parameters ($\vz, \theta, \psi$) and optimizer state to their previous best values as measured by the loss. The stage finishes after a certain number of steps or when the learning rate is decayed below a certain threshold value.
\end{description}

\subsubsection{Soft-rounding}
\label{app:sec:softrounding}

As discussed above, we warp the continuous latent values $\vz$ during the first stage of optimization with a soft-rounding function to better approximate the eventual quantization of the latents. 
A soft-rounding function is a differentiable relaxation of the (hard) rounding function; that is, it has a parameter $T$ (typically referred to as \emph{temperature}) whose value determines how well we approximate the hard rounding function. Crucially, by setting $T$ to a particular value ($T=0$ in our case), we recover the hard rounding function. As $T\rightarrow \infty$ our soft-rounding function (see below) tends to a linear function, equivalent to the straight-through gradient estimator that is used in \cc \cite{ladune2023cool}.
Note that despite using a soft-rounding function, we still have to add random noise to regularize the optimization. We explain this further in \cref{app:sec:kumaraswamy}.

Following~\citet{agustsson2020softrounddither}, we use a construction where we apply soft-rounding twice: first to the raw value $\vz$ and a second time after adding random noise $\vu$ to the result. That is, the $\mathrm{softround}_T(\vz, \vu)$ function in \cref{app:eq:method:stage1_objective,app:eq:method:stage1_net_grads,app:eq:method:stage1_z_grads} is given by
\begin{align}
    \label{app:eq:softrounding_final}
    \mathrm{softround}_T(\vz, \vu) & = r_T(s_T(\vz) + \vu),
\end{align}
where $r_T$ and $s_T$ are simple soft-rounding functions.

Following~\citet{agustsson2020softrounddither}, we used the following simple soft-rounding function,
\begin{align}
   s_T(z) &= \lfloor z \rceil + \frac{1}{2} \frac{\tanh(\Delta/T)}{\tanh(1/2T)} + \frac{1}{2}, \quad \Delta = z - \lfloor z \rfloor - \frac{1}{2},
\end{align}
which is invertible and differentiable everywhere. We further also use
\begin{align}
    r_T(y) &= s_T^{-1}(y - 0.5) + 0.5 \approx \mathbb{E}_X[X \mid s_T(X) + U = y]
\end{align}
for the second soft-rounding function (instead of applying $s_T$ again) as suggested by \citet{agustsson2020softrounddither}. Here, $X$ and $U$ are assumed to be uniform random variables. We found that $s_T(s_T(z) + u)$ seemed to perform equally well in our setting.

As the learning rate is decayed throughout stage 1, we can also decrease the temperature $T$ of the soft-rounding to better approximate the rounding operation. For simplicity we use a linear schedule that interpolates between a higher temperature ($T = 0.3$) at the beginning of stage 1 and a lower temperature ($T=0.1$) at the end of stage 1. A higher temperature corresponds to a more linear function while a lower temperature leads to a more step-like function, cf. \cref{fig:rouning_noise}.

\subsubsection{Kumaraswamy noise distribution}
\label{app:sec:kumaraswamy}

\begin{figure}[tb]
    \centering
    \includestandalone[width=\textwidth, mode=image]{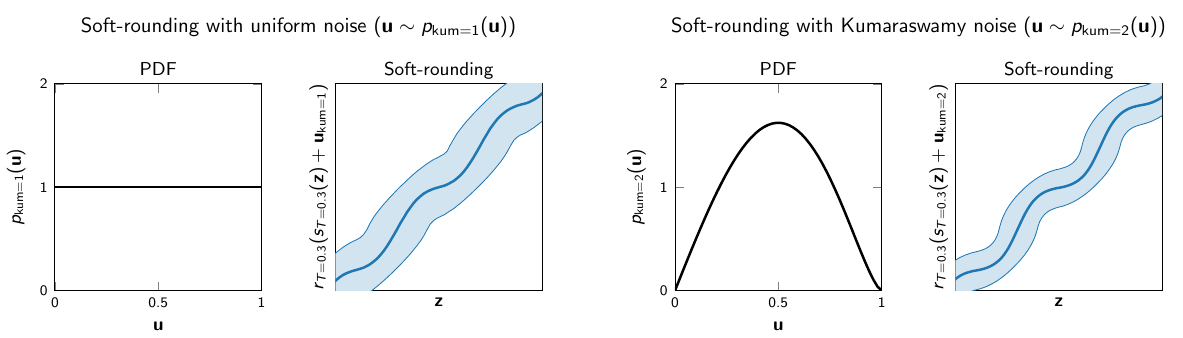}
    \caption{Probability density function (PDF) of the simplified Kumaraswamy distribution, cf. \protect\cref{app:eq:kumaraswamy_simplified}, and the effect of using it in the soft-rounding function (\protect\cref{app:eq:softrounding_final}) for two values of the shape parameter $a$. \textit{left:} $a=1$ corresponds to the uniform distribution; \textit{right:} $a=2$ yields a more peaked distribution that results in reduced variability of the soft-rounding function. For the soft-rounding we plot the mean and its $95$-percentiles.}
    \label{app:fig:kumaraswamy}
\end{figure}

As discussed in \cref{sec:method_opt}, adding noise during stage 1 is necessary even with soft-rounding because the (invertible) soft-rounding function alone does not create an information bottleneck. 
What do we mean by this? Hard rounding irreversibly destroys information by mapping all latent values in a certain quantization bin to the same (quantized) value.
Downstream computations, such as the reconstruction of the image with the synthesis network, then only have access to these quantized values. Therefore, it is important that the synthesis and entropy network are optimized in such a way as to only rely on the information in the quantized values, rather than information about the precise location of the latent within the quantization bin.
During optimization we use continuous valued latents as well as a continuous and invertible relaxation of rounding (as discussed in \cref{app:sec:softrounding}). And while the soft-rounding function can be steep for low temperatures $T$, its warping can be undone; that is, the synthesis network could learn to invert the soft-rounding function and then rely on information about location within the bin to improve the distortion loss without sacrificing the rate loss. The addition of noise is a mechanism to destroy this information in a \emph{differentiable} manner, such that we can still evaluate gradients of the objective, but prevent the networks from learning to use this information (that will get destroyed with quantization). 

A consequence of adding noise is that the gradients of the objective (\cref{app:eq:method:stage1_net_grads,app:eq:method:stage1_z_grads}) become stochastic, and the strength of the noise determines the variance of these gradients. Large gradient variance can lead to slower or worse optimization.
In particular when using the soft-rounding function, there is no reason \textit{a priori} that uniform noise should strike the best balance. We therefore explored other noise distributions as the detail in the following.

We want to flexibly parameterize the noise distribution in the compact interval $[0, 1]$. A natural choice for this is the $\Beta{a}{b}$-distribution that has two shape parameters $a$ and $b$ and can represent the Uniform distribution as well as symmetric and asymmetric overdispearsed (spread out) and underdispersed (peaked) distributions. However, we found that sampling from the $\mathrm{Beta}$-distribution is slow due to the transcendental functions involved in computing its density and CDF. We therefore use the Kumaraswamy distribution \cite{kumaraswamy1980} that is similar to the $\mathrm{Beta}$-distribution but has a closed form density and CDF.

The Kumaraswamy probability density function also has two shape parameters, $a$ and $b$, and is given by
\begin{align}
    p_{a, b}(u) & = abu^{a-1} (1-u^a)^{b-1}.
\end{align}

We are only interested in distributions with a mode at $0.5$ so as not to favor one direction; we can therefore simplify the distribution to only have a single parameter $a$:
\begin{align}
    \label{app:eq:kumaraswamy_simplified}
    p_{\text{kum}=a}(u)  & = (2^a (a-1) + 1) u^{a-1} (1-u^a)^{\tfrac{1}{a}(2^a-1)(a-1)}.
\end{align}
Setting $a=1$ corresponds to the uniform distribution, $p_{\text{kum}=1}(u) = \text{Uniform}(u)$.
We plot this simplified Kumaraswamy distribution (\cref{app:eq:kumaraswamy_simplified}) for $a=1$ and $a=2$ in \cref{app:fig:kumaraswamy}. Note that while the mode is at $u=0.5$, the distribution is not quite symmetric; yet we observed that this did not matter in practice, likely because the asymmetry is small.

In \cref{app:fig:kumaraswamy} we also show the effect of sampling from these distributions on the soft-rounding; as expected, sampling from a more peaked distribution, $p_{\text{kum}=2}$, leads to smaller uncertainty intervals at the same temperature ($T=0.3$ in this case) for the soft-rounding.

Because gradient variance is of more concern at high learning rates at the beginning of stage 1, the trade-off between regularization and optimization dynamics changes throughout stage 1. Empirically we found that linearly decaying the shape parameter $a$ from $a=2$ at the beginning of stage 1 to $a=1$ (uniform distribution) at the end of stage 1 performed best.

\subsubsection{Quantization and entropy encoding/decoding of the network parameters}
\label{app:sec:method:quantize_network_params}
Following \cc, we treat the synthesis and entropy parameters $\theta, \psi$ as continuous values during training, and quantize them separately after training. We do a grid search over the quantization steps for the weights and bias terms for $\theta$ and $\psi$ (so two terms in total, one for weight terms in either $\theta$ or $\psi$ and one for bias terms in either $\theta$ or $\psi$), that give quantized parameters $\widehat{\theta}$ and $\widehat{\psi}$. We select the quantization step that minimizes the following modified objective:
\begin{align}
    \mathcal{L}'_{\widehat{\theta}, \widehat{\psi}}(\vz) = 
    \|\vx - f_{\widehat{\theta}}\left(\mathrm{Upsample}(\vz)\right)\|_2^2 - \lambda (\log_2 P_{\widehat{\psi}}(\vz) + \log P(\widehat{\theta}) + \log P(\widehat{\psi}))
    \label{eq:quantization_step_search_objective}
\end{align}
Note that the RD-objective and the optimization thereof are not quantization-aware with respect to these network parameters. Addressing this may constitute interesting future work.

\subsection{Video: learning the custom masking}
\label{app:sec:learned_custom_masking}

\begin{figure}[ht]
    \small
    \centering
    \begin{subfigure}[b]{0.245\textwidth}
        \centering
        {Frame $0$}
        \includegraphics[width=\textwidth]{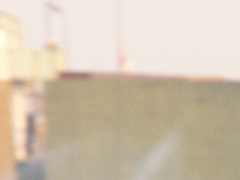}
    \end{subfigure}
    \begin{subfigure}[b]{0.245\textwidth}
        {Frame $1$}
        \centering
        \includegraphics[width=\textwidth]{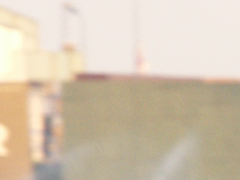}
    \end{subfigure}
    \begin{subfigure}[b]{0.245\textwidth}
        {Frame $2$}
        \centering
        \includegraphics[width=\textwidth]{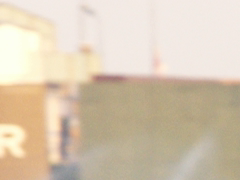}
    \end{subfigure}
    \begin{subfigure}[b]{0.245\textwidth}
        \centering
        {Frame $3$}
        \includegraphics[width=\textwidth]{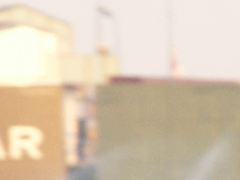}
    \end{subfigure}
    \begin{subfigure}[b]{0.245\textwidth}
        \centering
        \includegraphics[width=\textwidth]{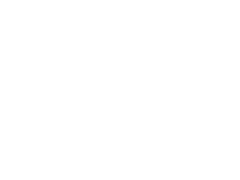}
        {}
    \end{subfigure}
    \begin{subfigure}[b]{0.245\textwidth}
        \centering
        \includegraphics[width=\textwidth]{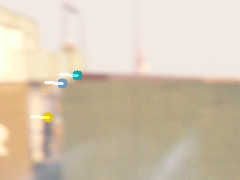}
        {Displacement $0 \rightarrow 1$}
    \end{subfigure}
    \begin{subfigure}[b]{0.245\textwidth}
        \centering
        \includegraphics[width=\textwidth]{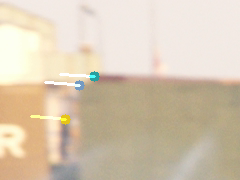}
        {Displacement $0 \rightarrow 2$}
    \end{subfigure}
    \begin{subfigure}[b]{0.245\textwidth}
        \centering
        \includegraphics[width=\textwidth]{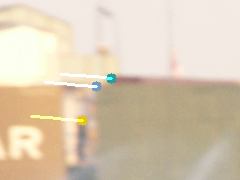}
        {Displacement $0 \rightarrow 3$}
    \end{subfigure}
    \caption{(Top) First few frames of a video patch from the Jockey sequence (UVG). (Bottom) displacement of key-points between consecutive frames computed using the OpenCV \cite{itseez2015opencv} implementation of Lucas-Kanade optical flow estimation \cite{lucas1981iterative}.}
    \label{app:fig:jockey}
\end{figure}

\begin{figure}[ht]
    \centering
    \includestandalone[mode=image]{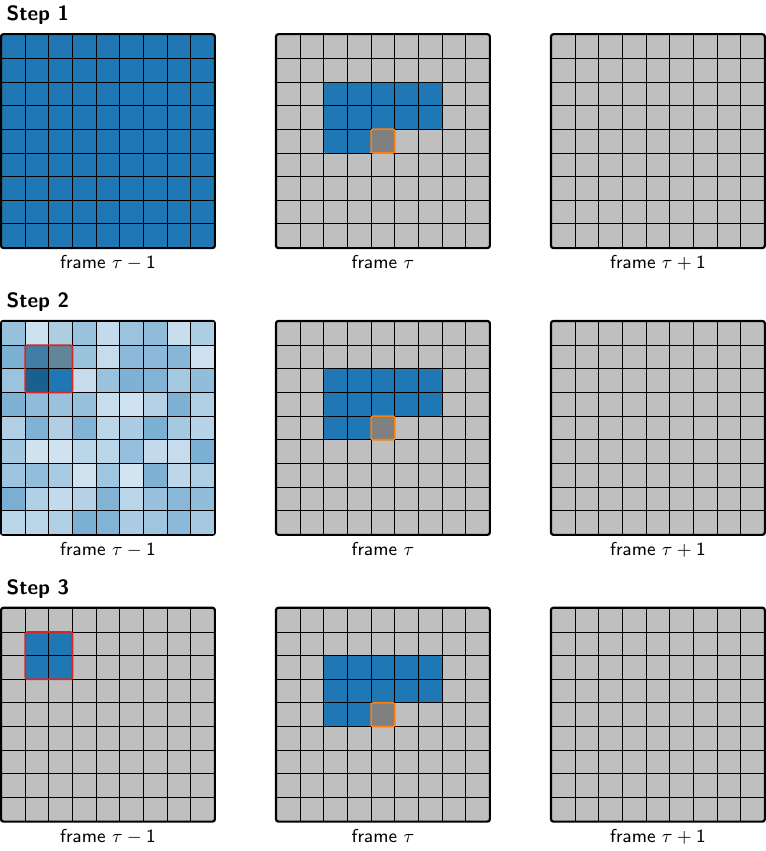}
    \caption{Visualization of the 3-step procedure for learning the custom masking.}
    \label{app:fig:learn_custom_mask_schematic}
\end{figure}

In the video setting, the context for predicting latent entries for a particular frame can also contain entries from the previous frame (cf. \cref{fig:video_context}). As discussed in \cref{sec:method_video}, it is important that the context is wide enough for the context of the previous frame to contain relevant information for predicting the latent entry of a particular frame.
For example, consider the first few frames of the video patch shown in the first row \cref{app:fig:jockey}. The second row shows the displacement of key-points between consecutive frames, and we can see that the displacement is greater than the small context width ($5-7$ latent pixels) that we use for images. In fact, the displacement of key-points between consecutive frames in the second row can be quantified using the Lucas-Kanade method for optical flow estimation \cite{lucas1981iterative}. This gives a mean displacement of $(19.5, 0.6)$ pixels per frame in the $x$ and $y$ direction respectively, where the mean is taken across the first 30 frames. Given that the latents and the synthesis network are designed in such a way that the latents only contain very local information about the video pixels, we would not be able to use the previous latent frame for predicting the current latent frame with a small context width. Hence we use a larger context width to be able to better capture motion.

The issue with na\"{i}vely using a larger context is that the number of entropy parameters grows with the context size, as we need $c_{\text{hidden}}$ entropy parameters for every context entry. Given that most of the latent entries in this wide context are irrelevant for predicting the target latent entry, we learn a custom mask for the previous latent frame context such that the entropy model can still access the relevant context in the previous latent frame while ignoring the irrelevant context therein.

Here we describe the procedure for learning this custom mask, with an overview in \cref{app:fig:learn_custom_mask_schematic}:
\begin{enumerate}
    \item \textbf{Train \method with wide spatial context for a few iterations.} First, train the entropy model with causal masking using a wide spatial context of size $ C\times C$ per latent grid for $M$ iterations, where $M$ is small. Typically we use $C=65$. Note that the wide context applies to both the previous latent frame $\tau - 1$ and the current latent frame $\tau$. We train with separate entropy parameters for each latent grid.
    \item \textbf{Compute magnitudes of entropy model weights for each context dimension.} For each latent grid, take the $C^2$ dimensions of the previous latent frame $\tau - 1$, and for each dimension compute the mean magnitude of the entropy model's first layer weights that process this dimension. \ie, suppose the entropy model's first $1 \times 1$ Conv layer for the previous latent frame context has weights of shape $C \times C \times 1 \times f_\text{out}$. Take the absolute value of these weights, followed by the mean across the final two axes to obtain the $C \times C$ magnitudes for each of the $C^2$ context dimensions.
    \item \textbf{Choose rectangular mask location that maximizes the sum of magnitudes within mask.} Given a fixed rectangular mask shape $c \times c'$ (where $c,c' \ll C$),
    sweep over all possible locations of the rectangular mask within the $C \times C$ context grid. For each location, compute the sum of the $c \times c'$ magnitudes within the mask. Then choose the location that has the highest sum.
\end{enumerate}
We thus obtain the $c \times c'$ learned rectangular mask for the previous latent frame $\tau -1$. We also empirically observed that for the lower resolution latent grids, there is little correlation between the latents for different frames, hence the entropy model doesn't use the previous latent frame for the prediction of the target latent pixel (orange border in \cref{app:fig:learn_custom_mask_schematic}). Hence we only use a learned mask for the $K$ highest resolution latent grids, and for the remaining grids we mask out all of the previous latent frame so that it is unused by the entropy model.

\begin{figure}[ht]
    \centering
    \includestandalone[mode=image]{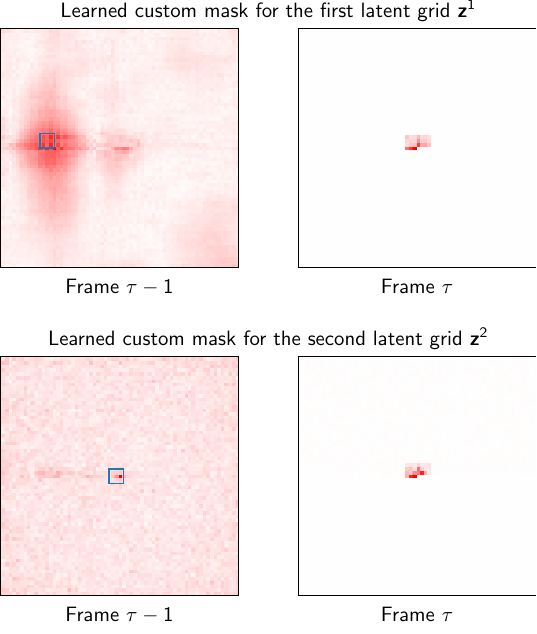}
    \caption{Visualization of the custom masks learned with the procedure described in \cref{app:sec:learned_custom_masking}. The heatmap represents the magnitude of the weights in step 2 of \cref{app:sec:learned_custom_masking}, and the blue box represents the learned custom masking for the previous latent frame for each latent grid.}
    \label{app:fig:custom_mask}
\end{figure}

Also note that for the current latent frame $\tau$, the relevant contexts for predicting the target latent pixel should only be a handful of values in the neighborhood of the target among the $(C^2 -1)/2$ causal context dimensions. So it would be a waste of entropy parameters (that need to be compressed and transmitted) to process the irrelevant context dimensions. 
Hence we use a small causal neighbourhood of size $l$ (where $l \leq C$ and $l$ is odd) around the target latent pixel, the same as for images \cref{app:fig:prev_grid_conditioning} (left).
Note that this causal neighbourhood for the current latent frame is fixed rather than learned, and used for all latent grids rather than just the $K$ highest resolution grids. Given this custom masking for the previous and latent frames, we train \method from scratch, fixing the custom mask. 

In practice we use $C=65,  M=1000, c=c'=4, K=3, l=7$; these values were obtained from a hyperparameter sweep on a small subset of patches of the UVG dataset.

In \cref{app:fig:custom_mask}, we show that we are able to learn a sensible custom mask with the above procedure when applied to a video patch of the Jockey sequence in \cref{app:fig:jockey}. We
use a low value of RD-weight $\lambda$ (\cref{eq:rate_distortion_objective}) for training, \ie, train for a high bitrate.

The heatmaps in \cref{app:fig:custom_mask} correspond to step 2 of the procedure above (\textit{compute magnitudes of entropy model weights for each context dimension}) when training on this video patch. After $M=1000$ iterations, we see that the entropy model for the first latent grid $\vz^1$ (top row) assigns the highest weights to the context dimensions that are consistent with the displacement calculated above, relative to the central target pixel. In step 3, the blue learned mask is placed here as this position has the greatest sum of magnitudes within the mask. For the second latent grid $\vz^2$ (bottom row), we see that the region corresponding to the displacement calculated above does indeed have higher weights than its neighborhood, but the highest weight is given to the central pixel. This indicates that the correlation between the latent dimensions that correspond to the same key-point in consecutive frames is weaker for $\vz^2$ compared to $\vz^1$. Given that we have trained for a high bitrate, most of the information content lies in $\vz^1$. This is consistent with the above observation that the aforementioned correlation is stronger for $\vz^1$ compared to $\vz^2$.

\begin{table}[htb]
    \centering
    \small
\begin{tabular}{lLc}
	\toprule
	\textbf{Hyperparameter} & \textbf{Fixed value} & \textbf{In \emph{adaptive} setting?}\\\midrule\midrule
	\multicolumn{3}{c}{Quantization -- Stage 1}  \\\midrulegray
	Number of encoding iterations & 10^5  &         \\
	\rowcolor{verylightgray} Initial learning rate & 10^{-2}          &      \\
	Final learning rate & 0     &      \\
	\rowcolor{verylightgray} Initial value of T for soft rounding & 0.3   &\\
	Final value of T for soft rounding & 0.1 &  \\
	\rowcolor{verylightgray} Initial value of $a$ for Kumaraswamy noise & \text{$2.0$ (\faCamera)\,\,/\,\, $1.75$ (\faVideoCamera)}   &  \\
	Final value of $a$ for Kumaraswamy noise & 1.0    &   \\
	\rowcolor{verylightgray} Threshold for gradient L2 norm clipping  &  
	\text{$10$ (\faCamera)\,\,/\,\, $0.03$ (\faVideoCamera)} &   \\ %
	\midrule %
	\multicolumn{3}{c}{Quantization -- Stage 2} \\\midrulegray
	Maximum number of encoding iterations & 10^4  & \\
	\rowcolor{verylightgray} Initial learning rate & 10^{-4} & \\
	Decay lr if loss has not improved for this many steps & 20 & \\
	\rowcolor{verylightgray} Decay lr by multiplying with this factor & 0.8 & \\
	Finish Stage 2 early if lr drops below this value & 10^{-8} &  \\
	\rowcolor{verylightgray} Value of $T$ for soft-rounding gradient estimation & 10^{-4} &  \\\midrule\midrule
	\multicolumn{3}{c}{ Architecture -- Latents}  \\\midrulegray
	Number of latent grids & - &  \\
	\rowcolor{verylightgray} Quantization step (bin width) for rounding & - &  \\
	Use the highest resolution grid $(\{t\}, h, w)$? & - & \faCamera{}  \\\midrule %
	\multicolumn{3}{c}{ Architecture -- Synthesis model}  \\\midrulegray
	Output channels of the $1\times 1$ convolutions (list) & - & \faCamera{} and \faVideoCamera  \\
	\rowcolor{verylightgray} Number of $3\times 3$ residual convolutions (with $3$ channels) & 2 &  \\
	Initialize the last layer bias with mean RGB of the image? & - &  \faVideoCamera \\
	\rowcolor{verylightgray} (video only) Replace $3\times3\times3$ Conv with $3\times3$ Convs per frame & - & \faVideoCamera \\\midrule %
	\multicolumn{3}{c}{ Architecture -- Entropy model} \\\midrulegray
	Widths of the $1\times 1$ convolutions & - & \faCamera{} and \faVideoCamera \\
	\rowcolor{verylightgray} Log-scale of Laplace is shifted by $\dots$ before $\exp$ & 8 &  \\
	Scale parameter of Laplace is clipped to & [10^{-3}, 150] &  \\
	\rowcolor{verylightgray} Context size (same grid) & - & \faCamera{} and \faVideoCamera \\
	Include previous grid in context? & - &  \faCamera \\
	\qquad Context size (previous grid) & 3 \times 3 &  \\\midrule
	\multicolumn{3}{c}{ Architecture -- Entropy model (video only)}  \\\midrulegray
	Learn separate models per grid, $\psi=(\psi_1, \dots \psi_N)$? & - & \faVideoCamera \\
	\rowcolor{verylightgray} Use custom masking (cf. \protect\cref{app:sec:learned_custom_masking})? & - &  \faVideoCamera \\
	\qquad  Mask size (current frame), $l$ & 7\times 7 &  \\
	\qquad Mask size (previous frame), $c \times c'$ & 4\times 4 &  \\
	\qquad Iteration count to learn the mask, $M$ & 1000 &  \\
	\qquad Number of grids for which mask is learned, $K$ & 3 &  \\\midrule %
	\multicolumn{3}{c}{ Other}  \\\midrulegray
	Possible quantization steps for network parameters & \clubsuit &  \\
	\rowcolor{verylightgray} (video only) Size of video patches & - &  \faVideoCamera \\\bottomrule
\end{tabular}
    \caption{Hyperparameters and their values for the quantization-aware optimization and architecture of \method for images (\faCamera) and video (\faVideoCamera). Where hyperparameters are fixed for all experiments and evaluations, their values are listed. Otherwise they are specified in the image or video specific hyperparameter list in \protect\cref{app:tab:method:hyperparameters_image,app:tab:method:hyperparameters_video}, respectively. An icon in the last column indicates whether a hyperparameter is included in the \textit{adaptive} setting for images and/or videos. Except for gradient L2 norm clipping, all quantization hyperparameter values are the same for images and videos. \qquad
    $\clubsuit$ The possible quantization steps for the network parameter $\theta$ and $\psi$ are $\{5 \cdot 10^{-5}, 10^{-4}, 5 \cdot 10^{-4}, 10^{-3}, 3 \cdot 10^{-3}, 6\cdot 10^{-3}, 10^{-2}\}$ for the weights and biases separately.}
    \label{app:tab:method:hyperparameters_opt_arch}
\end{table}

\subsection{Hyperparameters}
\label{app:sec:hyperparameters}

Here, we give an overview and a brief description of the hyperparameters used in our experiments as well as their settings. These comprise both architecture choices (and their hyperparameters) as well as optimization hyperparameters.
As explained in the main paper, for images we distinguish between evaluations with a single fixed setting for all images (that we simply denote as \method) and an adaptive setting where we select the best hyperparameter choice out of a small set on a per-image basis (we denote this as \method \emph{adaptive}). For videos we always select the best hyperparameters out of a small set on a per-patch basis.
The choices of varying hyperparameters form part of the header that is transmitted with the bitstream, as they are needed to decode the image.

In \cref{app:tab:method:hyperparameters_opt_arch} we provide a comprehensive list of all hyperparameters of \method. We also give their default values if they are fixed for all experiments and specify whether they are included in the \emph{adaptive} setting.

In \cref{app:tab:method:hyperparameters_image,app:tab:method:hyperparameters_video} we separately list all hyperparameters that differ for images and videos, respectively. We provide both fixed values as well as the possible sets of values that are explored in the \emph{adaptive} setting.

\paragraph{Adaptive setting for Kodak.} For Kodak, the \emph{adaptive} setting independently explores different values for three hyperparameters (see \cref{app:tab:method:hyperparameters_image}):
\begin{itemize}
    \item Whether the highest resolution latent grid is included ($2$ choices);
    \item Different entropy and synthesis network sizes ($3$ choices);
    \item Different context sizes for the entropy model ($2$ choices)
\end{itemize}
In total the \emph{adaptive} setting explores $2 \times 2 \times 3 = 12$ hyperparameter settings and picks the best one per image.

\paragraph{Adaptive setting for CLIC2020.} For CLIC2020, the \emph{adaptive} setting independently explores different values for two hyperparameters (see \cref{app:tab:method:hyperparameters_image}):
\begin{itemize}
    \item Whether the highest resolution latent grid is included ($2$ choices);
    \item Different entropy and synthesis network sizes ($3$ choices)
\end{itemize}
In total the \emph{adaptive} setting explores $2 \times 3 = 6$ hyperparameter settings and picks the best one per image.

\paragraph{Adaptive setting for UVG.} For UVG, we only use the \emph{adaptive} setting, which  explores different values for six hyperparameters that are grouped together as follows (see \cref{app:tab:method:hyperparameters_video}). See \cref{app:sec:video_ablations} for results using single/fewer settings. Namely we explore three "entropy parameter settings", \enticon{1}, \enticon{2}, \enticon{3}, that jointly specify the following hyper parameters:
\begin{itemize}
    \item Whether a separate entropy model is learned per grid;
    \item Different context sizes for the entropy model;
    \item Whether custom masking is used
\end{itemize}
Similarly, we explore three "synthesis parameter settings", \synthicon{1}, \synthicon{2}, \synthicon{3}, that jointly specify the following hyperparameters:
\begin{itemize}
    \item Whether the bias of the last layer of the synthesis network is initialized to the mean RGB values of the image;
    \item Whether the $3\times 3\times 3$ 3D convolutions in the synthesis model are replaced by $3\times 3$ 2D convolutions per frame.
\end{itemize}
In total the adaptive setting explores $3 \times 3 = 9$ hyperparameter settings and picks the best one per video patch.

Moreover, the video patch size is chosen according to the RD-weight $\lambda$, as we observed that larger patches give better RD performance for high values of $\lambda$ (low bitrates) and vice versa. We use $(30 \times 180 \times 240)$ for $\lambda \leq 2\cdot 10^{-4}$, $(60 \times 180 \times 240)$ for $2\cdot 10^{-4} < \lambda \leq 10^{-3}$, $(75 \times 270 \times 320)$ for $\lambda > 10^{-3}$.

\begin{table}[htb]
    \centering
\begin{tabular}{lCC}
	\toprule
	\textbf{Hyperparameter} & \textbf{Fixed value} & \textbf{Adaptive values}\\\midrule\midrule
	Number of latent grids & 7 & - \\
	\rowcolor{verylightgray} Latent quantization step (bin width) for rounding & 0.4 & - \\
	Use the highest resolution grid $(h, w)$? & \text{\cmark} & \text{\cmark, \xmark} \\ \midrulegray
	\rowcolor{verylightgray} Output channels of synthesis $1 \times 1$ convs${}^{\clubsuit}$ & (18, 18) & (12, 12), (18, 18), (24, 24) \\
	Initialize the last synthesis layer bias with mean RGB of the image? & \text{\xmark} & - \\ \midrulegray
	\rowcolor{verylightgray} Output channels of entropy $1 \times 1$ convs${}^{\clubsuit}$ & (18, 18) & (12, 12), (18, 18), (24, 24) \\
	Learn separate entropy model per grid, $\psi=(\psi_1, \dots \psi_N)$?  & \text{\xmark} & - \\\midrule\midrule
	\multicolumn{3}{c}{ Kodak only}  \\ \midrulegray
	Include previous grid in context? & \text{\xmark} & - \\
	\rowcolor{verylightgray} Context size (same grid) in entropy model & 7\times 7 & 5\times 5, 7 \times 7 \\ \midrule\midrule
	\multicolumn{3}{c}{ CLIC2020 only}  \\\midrulegray
	Include previous grid in context?& \text{\cmark} & - \\
	\rowcolor{verylightgray} Context size (same grid) in entropy model & 7\times 7 & -\\ \bottomrule
    \end{tabular}
    \caption{Hyperparameter values that are specific to images. \textit{Fixed value} contains the values that are fixed for all images. It also contains the default values in the (non-adaptive) setting. \textit{Adaptives values} specifies the values that are explored in the adaptive setting.\\
    ${}^{\clubsuit}$ The adaptive values for the synthesis and entropy model sizes are varied together.
    }
    \label{app:tab:method:hyperparameters_image}
\end{table}

\begin{landscape}
\begin{table}
    \centering
    \begin{tabular}{l C MMM MMM}
	\toprule
	\multirow{2}{*}{\textbf{Hyperparameter}} &\multirow{2}{*}{\textbf{Fixed value}} & \multicolumn{6}{c}{\textbf{Adaptive values}}\\
	&& \enticon{1}& \enticon{2}&\enticon{3}& \synthicon{1}&\synthicon{2}&\synthicon{3}\\\midrule\midrule
	Size of video patches & - & \multicolumn{6}{C}{(30 \times 180 \times 240), (60 \times 180 \times 240), (75 \times 270 \times 320)^{\clubsuit}} \\
	\rowcolor{verylightgray} Number of latent grids & - & 6 & 5 & 6 &&& \\
	Latent quantization step (bin width) for rounding & 0.3 & &&&&& \\
	\rowcolor{verylightgray} Use the highest resolution grid $(t, h, w)$? & \text{\cmark} & &&&&& \\ \midrulegray
	Learn separate entropy model per grid, $\psi=(\psi_1, \dots \psi_N)$? & - & \text{\xmark} & \text{\cmark} & \text{\cmark}  &&\\
	\rowcolor{verylightgray} Output channels of entropy $1 \times 1$ convs & - & (16, 16) & (2, 2) & (8, 8) &&&  \\
	Context size (same grid) in entropy model & - & {\small 3\times 9\times 9} & {\small 3\times 9\times 9} & {\small 3\times 65 \times 65} &&& \\
	\rowcolor{verylightgray} Use custom masking? & - & \text{\xmark}  & \text{\xmark}  & \text{\cmark} &&& \\
	Include previous grid in context? & \text{\xmark}&  &&&&&\\\midrulegray
	\rowcolor{verylightgray} Output channels of synthesis $1 \times 1$ convs & (32, 32) & &&&&& \\
	Initialize the last synthesis bias with mean RGB of the image? & - &&&& \text{\cmark} & \text{\cmark}& \text{\xmark}\\ 
	\rowcolor{verylightgray} Replace $3\times3\times3$ Conv with $3\times3$ Convs per frame & - &&&& \text{\cmark} & \text{\xmark}& \text{\cmark}\\\bottomrule
    \end{tabular}
    \caption{Hyperparameter settings that are specific to videos. \textit{Fixed value} contains the hyperparameter values that are fixed for all images. \textit{Adaptive values} lists the hyperparameter values that are explored for each patch separately. There are three possible settings for the entropy model, \protect\enticon{1}, \protect\enticon{2}, \protect\enticon{3}, and three possible settings for the synthesis model, \protect\synthicon{1}, \protect\synthicon{2}, \protect\synthicon{3}. Therefore, for each patch, we explore $3 \times 3 = 9$ different settings.\\
    ${}^\clubsuit$ The video patch size is chosen according to the RD-weight $\lambda$. We use $(30 \times 180 \times 240)$ for $\lambda \leq 2\cdot 10^{-4}$, $(60 \times 180 \times 240)$ for $2\cdot 10^{-4} < \lambda \leq 10^{-3}$, $(75 \times 270 \times 320)$ for $\lambda > 10^{-3}$.}
    \label{app:tab:method:hyperparameters_video}
\end{table}
\end{landscape}

\clearpage
\section{Evaluation details} 
\label{app:sec:evaluation_details}

\subsection{BD-rate}
The Bj{\o}ntegaard Delta rate (BD-rate) metric \cite{bjontegaard2001calculation} is a scalar that estimates the saving in bitrate of one RD curve compared to another. It is useful since it allows to quantify the improvement of one RD curve over another with a single scalar.
Given an anchor RD curve and a candidate RD curve, the curves are first transformed into a curve of distortion vs log-rate. Then the difference between the area under the curve (with respect to the distortion/PSNR axis) of the candidate and the anchor are computed. Note that since the area under the curve is measured with respect to the distortion axis, the smaller the area under the curve, the better. Therefore a negative value of the BD-rate implies that the candidate curve is better than the anchor.
Also note that the output is invariant to (positive) scalar multiplication of the rate, due to the use of the log rate. Hence the BD rate should be invariant to whether we use bpp, bits or nats.

\subsection{PSNR evaluation for videos}
We compute PSNR with the following convention also used in \citet{mentzer2022vct}: take the per-frame PSNR for each frame of a given video, then take the mean across all frames for that video to get a PSNR value for that video. Take the mean of these values across all videos to get the PSNR for a given RD-weight. For bpp, simply take the mean across all patches of a video to get the bpp for a given video, then take the mean across all videos.

\subsection{Entropy coding}

\begin{figure}[ht]
    \centering
    \includestandalone[mode=image]{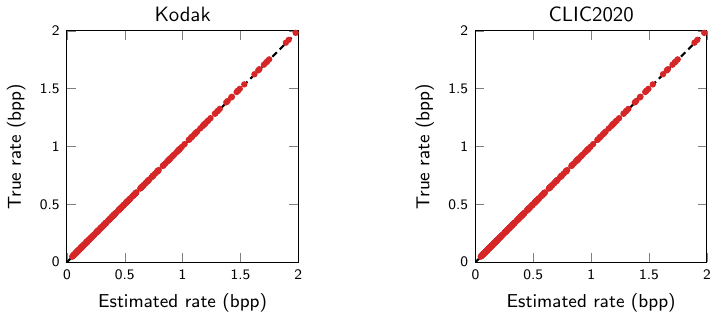}
    \caption{The true bit-rate observed when range coding images into a bit-stream is nearly identical to our estimates of the bit-rate obtained by evaluating log-probabilities. Each point corresponds to an image encoded at a given rate-distortion weight.}
    \label{app:fig:true_bpp}
\end{figure}

To encode images to files, we use the \texttt{range-coder} Python package available in PyPI. First, hyperparameter choices and other information that depend on the image are uniformly encoded. These currently include the image resolution and the choice of quantization step widths for the synthesis transform and the entropy model. Other hyperparameters (such as the model architecture) are assumed to be fixed and known to the decoder. Next, the weights and biases of the synthesis transform and entropy model are encoded assuming a different zero-mean Laplace distribution for each layer's set of weights and set of biases. The scale of the Laplace distribution is estimated from the parameters themselves, quantized to one of 1024 possible values, and uniformly entropy encoded using the range coder. Finally, the latent grids are autoregressively encoded with a Laplace distribution whose mean and scale are predicted by the entropy model.

For convenience and faster turnaround times, bpp values used in plots throughout the paper were estimated by evaluating the log-probabilities that would be used when range coding model parameters and latent values. We find that these estimates are very close to the bit-rates observed when range coding using the default, non-adaptive setting of C3 (Fig.~\ref{app:fig:true_bpp}).

\subsection{Estimating the decoding complexity in MACs}

Here we explain in more detail how we obtained the estimates for the decoding complexity. We report this value in terms of the number of multiply-accumulate operations (MACs) per pixel. We follow the convention that $1\,\text{MAC} = 2\,\text{FLOPs}$, though note that this is not always consistently done in the literature.

We follow \cc \cite{ladune2023cool} and \cctwo \cite{leguay2023low} and report theoretic MACs for applying the neural networks in our model, \ie, matrix multiplications, but exclude pointwise operations such as non-linearities. We also include an estimate for bi-/trilinear upsampling.

\cctwo uses \fvcore \cite{fvcore} to automatically estimate the number of theoretic MACs in \pytorch \cite{pytorch}. We implement our method in \jax \cite{jax}, which (to the best of our knowledge) does not easily allow for the automatic estimation of theoretic MACs/FLOPs. We therefore use back-of-the-envelope estimates as detailed below and confirm they agree with the numbers as reported by \fvcore in a \pytorch codebase.

The MACs estimates for other baselines were obtained from various sources while ensuring the numbers were comparable with ours. For BMS \cite{balle2018variational}, MBT \cite{minnen2018joint} and CST \cite{cheng2020learned}, we calculated the MACs using the \fvcore library on the CompressAI implementations of these models (measuring MACs only on the decoder of the model).
For MLIC and MLIC+ \cite{jiang2022multi}, the numbers were directly provided to us by the authors, who calculated their numbers using the \deepspeed library. For EVC \cite{wang2023evc}, we used the numbers for the EVC-S, EVC-M and EVC-L models in the paper, which were obtained using the \ptflops library.

\paragraph{Bi- and trilinear upsampling}

To upsample the latent grids we use bilinear upsampling for images and trilinear upsampling for videos, respectively.
\fvcore does not provide a complexity estimate for upsampling; we therefore use the following upper-bound estimates.
We upper bound the complexity of bilinear upsampling with $8$ MACs per output pixel; this estimate includes computing the weighted average of the values at the four closest grid points ($4$ MACs) and computation of the corresponding weights ($4$ MACs).
Similarly, we upper bound the complexity of trilinear upsampling with $16$ MACs per output pixel; this estimate includes computing the weighted average of the values at the eight closest grid points ($8$ MACs) and computation of the corresponding weights ($8$ MACs).
Note that in practice, most of the weights can be pre-computed and cached, especially when we are upsampling by an exact factor of $2$.

\paragraph{Application of the entropy model.} While we implement the entropy as a convolutional network with a masked convolution as first layer and a sequence of $1\times 1$ convolutions during encoding, it can be equivalently evaluated as a feed-forward MLP where the context for each latent grid location is read from memory. Each latent grid location is evaluated independently in this case.
Therefore, the cost of evaluating the entropy model on all latent entries is given by the cost of applying it to one entry multiplied by the number of latent values $\mathcal{Z}$, which is given by 
\begin{align}
    \mathcal{Z} = \sum_n h_n \cdot w_n
\end{align}
where $h_n$ and $w_n$ are the height and width of latent grid $n$.
The cost of applying the entropy model is the cost of applying all layers, which depends on the input, hidden, and output sizes. The cost in MACs of each single layer is simply the product of the input and output size, $c_\text{in} \cdot c_\text{out}$.
The total complexity in MACs/pixel is then given by adding the contributions from all latent values and dividing them by the number of pixels.

We verify that the numbers we obtain for the application of the feed-forward MLP are the same as those reported by \fvcore that is used by \cctwo.
When estimating the cost of additionally conditioning on the previous grid, we need to take into account the larger input-size to the entropy network as well as the bilinear upsampling of the latent. We upper-bound the cost of the bilinear upsampling by $\mathcal{Z} \cdot \mathrm{cost}_\text{one upsampling}$ as we have to resample exactly one value from a previous grid for each current grid location. This is an upper bound because the first grid does not actually depend on a previous grid.

\paragraph{Application of the synthesis model}
The synthesis model \synthnet is a simple convolutional network with skip connections in its second part. The cost of applying a single convolutional layer at one location is
\begin{align}
    k \cdot k \cdot c_\mathrm{in} \cdot c_\mathrm{out}.
\end{align}
As the number of locations is given by the number of pixels, the above estimate is the cost of applying a single layer in MACs/pixel.
We again evaluate the cost for different network sizes and verify that they agree with the numbers reported by \fvcore when implementing them.

\begin{table}[htb]
    \centering
    \begin{tabular}{lCCCC}
    \toprule
    \textbf{Component}& \textbf{KODAK}  & \textbf{CLIC} & \textbf{UVG} \\\midrule
    Entropy model & 1600 & 1889 & 2540\\
    Upsampling    & 48 & 48 & 80\\
    Synthesis model & 978 & 978 & 1798\\\midrule
    Total & 2626 & 2925 & 4418\\\bottomrule
    \end{tabular}
    \caption{Maximum computational complexity in MACs/pixel of different components of \method for the hyperparameters used in each dataset.}
    \label{app:tab:theoretic_macs}
\end{table}
\section{Additional results}
\label{app:sec:additional_results}

\subsection{Full RD curves with all baselines}
\label{app:sec:full_rd_curves}

\begin{figure}[tb]
  \centering
  \includestandalone[mode=image]{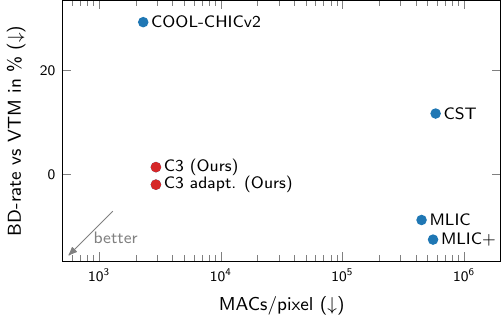}
   \caption{MACs/pixel vs BD-rate on CLIC2020. \method performs well both in terms of BD-rate and decoding MACs/pixel, achieving a better trade off than existing neural codecs.}
   \label{app:fig:bd-rate-vs-macs-clic}
\end{figure}

{
\begin{figure}[p]
  \centering
  \includestandalone[mode=image]{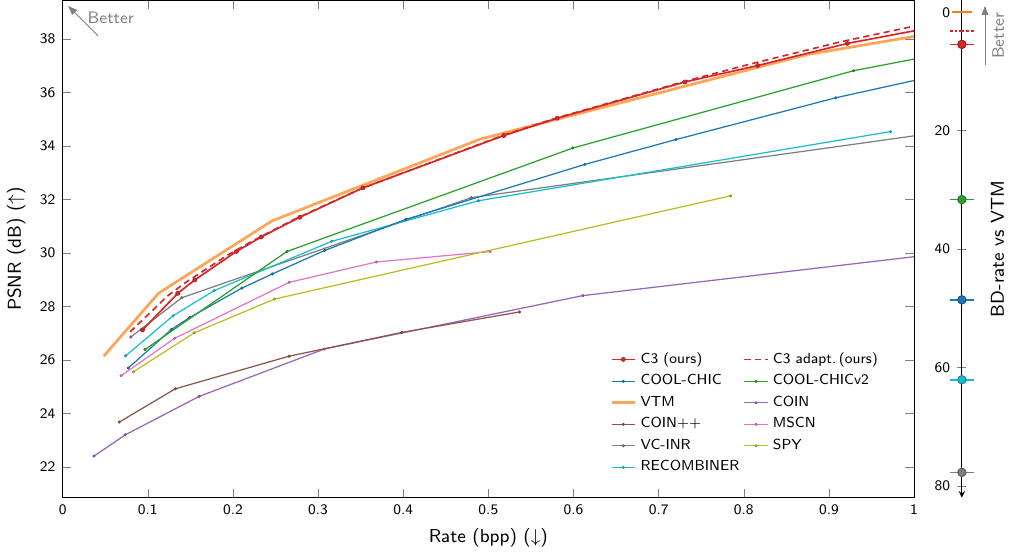}
   \caption{Rate-distortion curve and BD-rate on the Kodak image benchmark comparing \method to other overfitted neural field based compression methods, including those in \protect\cref{fig:rd-curve-kodak}. Note that we omit methods with very large values from the BD-rate plot on the right.}
   \label{app:fig:rd-curve-kodak-single-image}
\end{figure}

\begin{figure}[p]
  \centering
  \includestandalone[mode=image]{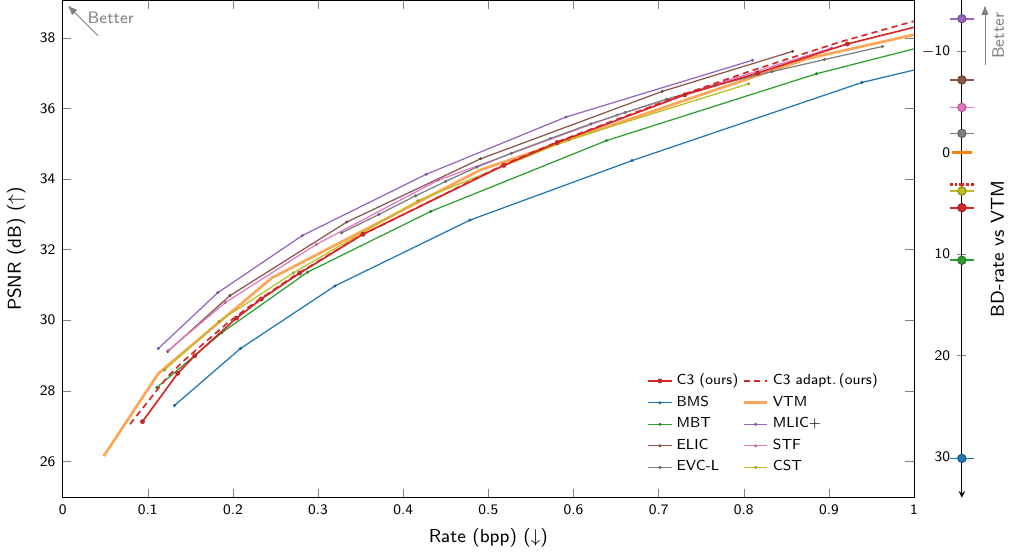}
   \caption{Rate-distortion curve and BD-rates on the Kodak image benchmark comparing \method to autoencoder-based neural compression methods, including those in \protect\cref{fig:rd-curve-kodak}.}
   \label{app:fig:rd-curve-kodak-autoencoders}
\end{figure}

\begin{figure}[p]
  \centering
  \includestandalone[mode=image]{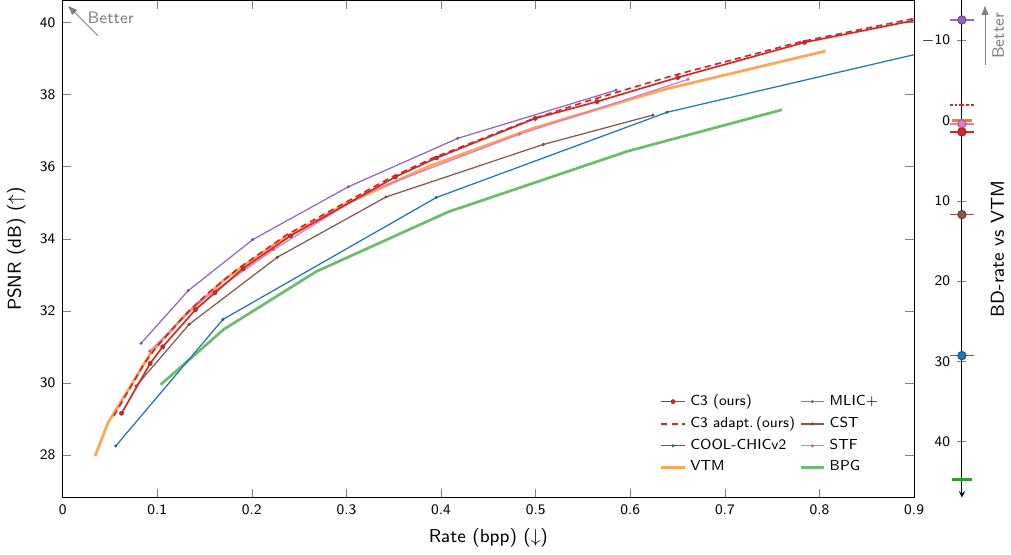}
   \caption{Rate-distortion curve and BD-rates of more baselines on CLIC2020, including those in \cref{fig:rd-curve-clic}.}
   \label{app:fig:rd-curve-clic-full}
\end{figure}

\begin{figure}[p]
  \centering
  \includestandalone[mode=image]{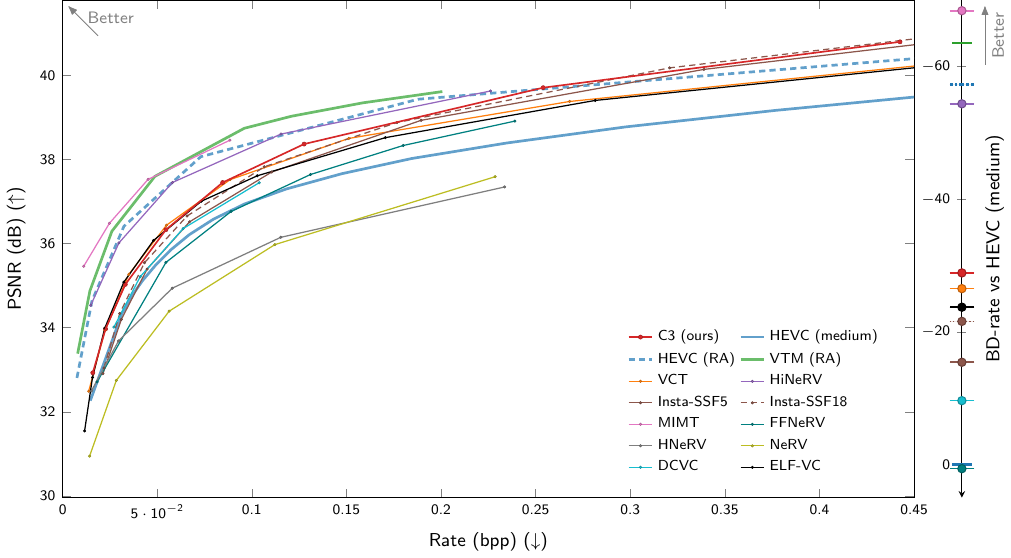}
   \caption{Rate-distortion curve of more baselines on all UVG videos, including those in \cref{fig:rd-curve-uvg}.}
   \label{app:fig:rd-curve-uvg-full}
\end{figure}

\begin{figure}[p]
    \centering
    \includestandalone[mode=image]{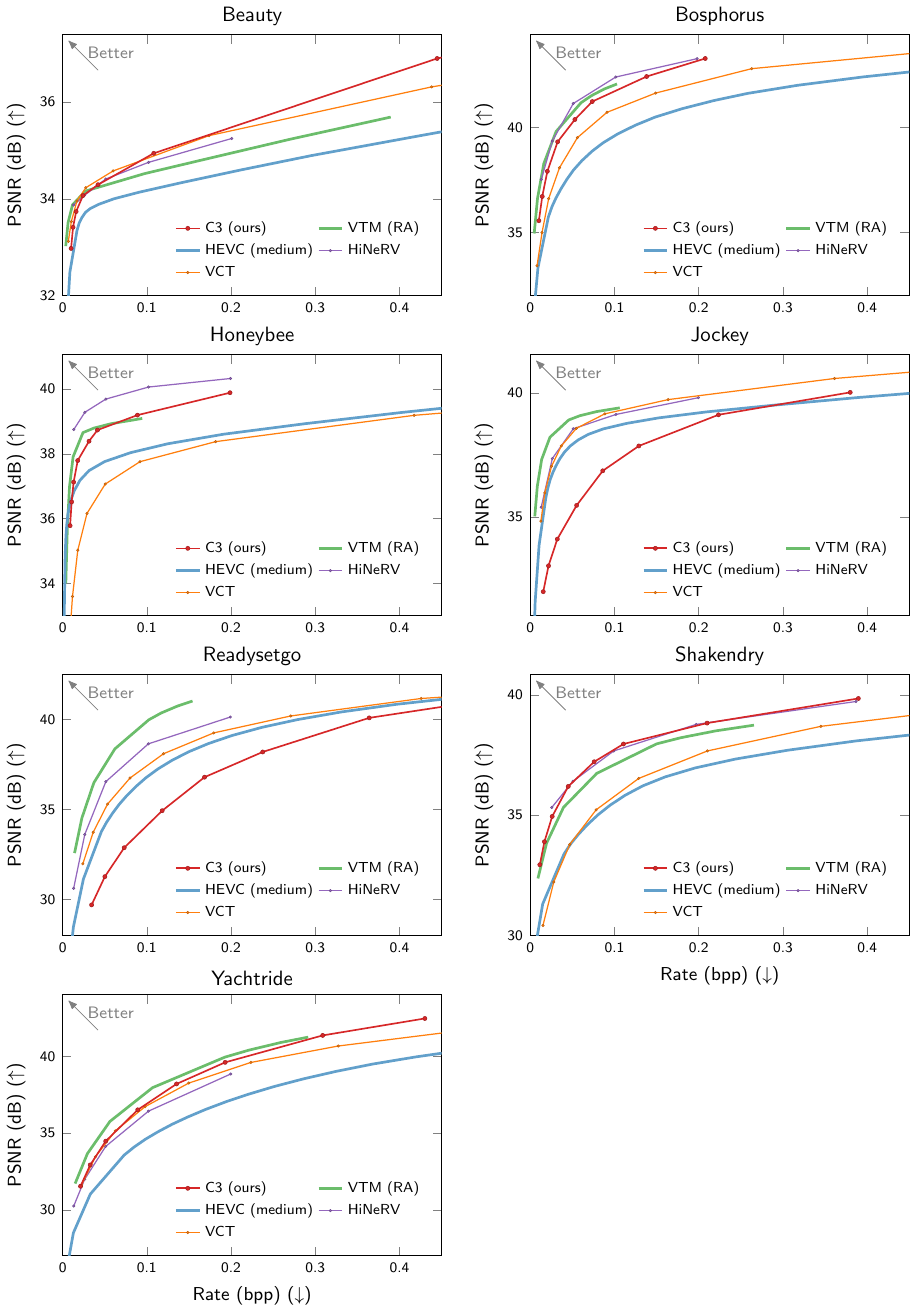}
    \caption{Rate-distortion curves for individual videos in UVG.}
    \label{app:fig:rd-curve-uvg-individual}
\end{figure}
}

We include several large-format RD-curve plots in which we compare \method to:
\begin{enumerate}
    \item overfitted neural compression methods on the Kodak image benchmark in \cref{app:fig:rd-curve-kodak-single-image},
    \item autoencoder based neural compression methods on the Kodak image benchmark in \cref{app:fig:rd-curve-kodak-autoencoders},
    \item several additional baselines on the CLIC2020 image benchmark in \cref{app:fig:rd-curve-clic-full}, and
    \item several additional baselines on the UVG video benchmark in \cref{app:fig:rd-curve-uvg-full}.
\end{enumerate}

\paragraph{Image baselines.} In addition to the baselines used in the main paper, for images we also compare against\footnote{We follow the CompressAI \cite{begaint2020compressai} convention of using the first letters of the first three authors for unnamed methods}: COIN \cite{dupont2021coin}, SPY \cite{strumpler2022implicit}, COIN++ \cite{dupont2022coin++}, MSCN \cite{schwarz2022meta}, VC-INR \cite{schwarz2023modality}, RECOMBINER \cite{he2023recombiner}, MBT \cite{minnen2018joint}, ELIC \cite{he2022elic} and STF \cite{zou2022devil}. COOL-CHIC, COOL-CHICv2, COIN, COIN++ and STF results were obtained from official code implementations. MLIC, SPY, MSCN, VC-INR and RECOMBINER results were obtained from direct communication with the respective paper authors. BPG, VTM, BMS, MBT, CST and ELIC were obtained from CompressAI \cite{begaint2020compressai}. 

\paragraph{Video baselines.} In addition to the baselines cited in the main paper, for videos we also include HEVC (HM 18.0, Random Access, default setting) \cite{sullivan2012overview}, VTM (17.0, Random Access, default setting) \cite{bross2021overview}, Insta-SSF18 \cite{van2021instance} (bigger model than Insta-SSF5), DCVC \cite{li2021deep}, ELF-VC \cite{rippel2021elf}, NeRV \cite{chen2021nerv} and HNeRV \cite{chen2023hnerv}. HEVC (RA), NeRV, HNeRV, FFNeRV and HiNeRV results were obtained from \citet{kwan2023hinerv} or direct communication with the authors. HEVC (medium, no B-frames), DCVC, ELF-VC and VCT results were obtained from direct communication with the authors of \citet{mentzer2022vct}. Insta-SSF5/18 and MIMT results were obtained directly from their papers. VTM (RA) results were obtained by running VTM 17.0 using the default setting of \texttt{encoder\_randomaccess\_vtm.cfg} \cite{VVenC}.

\paragraph{Discussion of the additional results.}
From \cref{app:fig:rd-curve-kodak-single-image}, we see that on the Kodak image benchmark, \method is SOTA among methods based on overfitting neural fields to a single image instance by a noticeable margin, and is the only method that is competitive with VTM . From \cref{app:fig:rd-curve-kodak-autoencoders}, we see that \method is not as competitive as the more recent autoencoder-based methods in terms of RD, although the gap has been significantly reduced compared to \cctwo. 

On CLIC2020, arguably a more realistic dataset than Kodak with images at higher resolution, we see in \cref{app:fig:rd-curve-clic-full} that \method \emph{adaptive} outperforms VTM and is quite close to MLIC+, the best-performing baseline. We also show the plot of BD-rate against decoding complexity in \cref{app:fig:bd-rate-vs-macs-clic}; 
\method performs much better in terms of RD performance compared to other baselines with a low decoding complexity such as COOL-CHICv2.

On UVG, we show in \cref{app:fig:rd-curve-uvg-full} that \method is also a competitive baseline for video compression, despite being the first one of its kind (in the COOL-CHIC line of work) to be applied to videos.
Note that there is a clear room for improvement that is visible in the gap between \method and stronger baselines such as HiNeRV and MIMT, however we emphasize again that \method achieves its competitive performance with 2-3 orders of magnitude lower decoding complexity than these baselines (cf. \cref{fig:bd-rate-vs-macs-uvg}).

\subsection{Encoding times}
See \cref{app:tab:encoding_times} for details on encoding times for each dataset, showing the fastest and slowest settings %
among the hyperparameter settings in the adaptive sweeps. Note that the encoding time for CLIC2020 depends on the size of the image, hence we measure it on the largest image of resolution $1370 \times 2048$. We emphasize again that we do not optimize for encoding times and use unoptimized research code to obtain these encoding times.

\begin{table}[h!]
    \centering
    \small
\begin{tabular}{lLC}
	\toprule
	\textbf{Hyperparameter} & \textbf{Value} & \textbf{Encoding time (sec/1k steps)}\\\midrule\midrule
	\multicolumn{2}{c}{\small Kodak -- fastest setting} & 3.9 \\\midrulegray
	Context size (same grid) & 5 \times 5 &  \\
	Width of $1\times 1$ convolutions (synthesis \&  entropy) & (12, 12) &  \\
	Use the highest resolution grid $({t}, h, w)$? & \text{\xmark} &  \\
	\midrule
	\multicolumn{2}{c}{\small Kodak -- slowest setting} & 7.1 \\\midrulegray
	Context size (same grid) & 7 \times 7 &  \\
	Width of $1\times 1$ convolutions (synthesis \&  entropy) & (24, 24) &  \\
	Use the highest resolution grid $({t}, h, w)$? & \text{\cmark} &  \\
	\midrule\midrule
	\multicolumn{2}{c}{\small CLIC2020 -- fastest setting} & 21.5 \\\midrulegray
	Width of $1\times 1$ convolutions (synthesis \&  entropy) & (12, 12) &  \\
	Use the highest resolution grid $({t}, h, w)$? & \text{\xmark} &  \\
	\midrule
	\multicolumn{2}{c}{\small CLIC2020 -- slowest setting} & 48.0 \\\midrulegray
	Width of $1\times 1$ convolutions (synthesis \&  entropy) & (24, 24) &  \\
	Use the highest resolution grid $({t}, h, w)$? & \text{\cmark} &  \\
	\midrule\midrule
	\multicolumn{2}{c}{\small UVG -- fastest setting} & 28.7 \\\midrulegray
	Patch size & (30, 180, 240) &  \\
	Entropy setting & \text{No conditioning} &  \\
	Replace $3\times3\times3$ Conv with $3\times3$ Convs per frame & \text{\cmark} &  \\
	\midrule
	\multicolumn{2}{c}{\small UVG -- slowest setting} & 456.7 \\\midrulegray
	Patch size & (75, 270, 320) &  \\
	Entropy setting & \text{Learned mask} &  \\
	Replace $3\times3\times3$ Conv with $3\times3$ Convs per frame & \text{\xmark} & \\\bottomrule
\end{tabular}
    \caption{Encoding times for \method measured on a single NVIDIA V100 GPU.}
    \label{app:tab:encoding_times}
\end{table}

\subsection{RD curves for individual UVG videos}
In \cref{app:fig:rd-curve-uvg-individual}, we include RD curves for individual UVG videos. Note that \method tends to perform better than VCT for Beauty, Bosphorus, Honeybee, Shakendry and Yachtride, and even outperforms VTM (17.0, random access setting) on Beauty and Shakendry. However \method appears to struggle with Jockey and Readysetgo, which are the video sequences with faster motion. 
While we show in \cref{app:sec:video_ablations} that the learned mask helps to achieve a better RD performance, it would be interesting to investigate how the performance can be improved further especially on these sequences with fast motion.

\clearpage
\section{Additional ablations}
\label{app:sec:additional_ablations}

\subsection{CLIC2020 ablations}
\label{app:sec:clic_ablations}

In \cref{app:tab:ablations_clic} we show the ablations for CLIC2020 when sequentially removing each of our improvements from the best performing \method Adpative model, similarly to \cref{tab:ablations} for Kodak in the main paper. In \cref{app:tab:ablation_knockout_clic}, we show the results when disabling individual features from the \method model, similar to \cref{tab:ablation_knockout} for Kodak in the main paper. We have an additional row for the previous grid conditioning described in \cref{app:sec:entropy_model_details}, since this option is only used for CLIC2020. The conclusions are similar in that soft-rounding, the GELU activation function and Kumaraswamy noise account for most of the boost in RD performance. Two minor differences for CLIC2020 are: 1) the quantization step is slightly more important than the remaining features compared to Kodak, 2) conditioning on the previous grid improves performance for CLIC2020. In \cref{app:tab:ablation_annealing} we ablate the annealing schedules for the soft-rounding temperature and the shape parameter of the Kumaraswamy noise. We find that fixing them to a single value leads to worse performance. A lower soft-rounding temperature results in a closer approximation of quantization but higher variance in the gradients. Annealing the temperature allows us to control this bias-variance trade-off over time, with less biased gradients becoming more important later in training. Also note that different shape parameters of the noise are optimal for different soft-rounding temperatures.

\begin{table}[htb]
    \centering
    \includestandalone[mode=image]{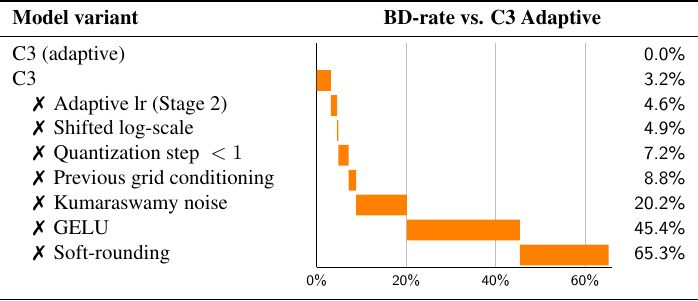}
  \caption{CLIC2020 ablation sequentially removing methodological changes. Note that a higher BD-rate means worse RD performance relative to our default settings for C3.}
  \label{app:tab:ablations_clic}
\end{table}

\begin{table}[htb]
    \centering
    \begin{tabular}{lC}
    \toprule
    \textbf{Removed Feature} & \textbf{BD-rate vs. \method} \\
    \midrule
    C3 $-$ Soft-rounding & 24.87\% \\ %
    C3 $-$ GELU & 9.83\% \\ %
    C3 $-$ Kumaraswamy noise & 8.06\%\\ %
    C3 $-$ Quantization step ${}< 1$ & 2.82\% \\  %
    C3 ${}-$ Previous grid conditioning & 1.50\% \\ %
    C3 $-$ Shifted log-scale  ${}> 0$ & 0.99\% \\ %
    C3 $-$ Adaptive lr (Stage 2) & 0.54\% \\ %
    \bottomrule
    \end{tabular}
    \caption{CLIC2020 ablation knocking out individual features from \method (fixed hyperparameters across all images). Note that a higher BD-rate means worse RD performance relative to our default settings for C3.}
    \label{app:tab:ablation_knockout_clic}
\end{table}

\begin{table}[htb]
    \centering
    \begin{tabular}{CCC}
    \toprule
    \textbf{Soft-round temperature} \hspace{1mm} \emph{T} & \textbf{Kumaraswamy noise} \hspace{1mm} \emph{a} & \textbf{BD-rate vs. \method} \\
    \midrule
    0.1  & 1.0 &  88.48\% \\
    0.1  & 1.5 & 109.08\% \\
    0.1  & 2.0 & 129.96\% \\
    0.2  & 1.0 &  11.66\% \\
    0.2  & 1.5 &   5.66\% \\
    0.2  & 2.0 &   5.97\% \\
    0.3  & 1.0 &  10.32\% \\
    0.3  & 1.5 &   5.57\% \\
    0.3  & 2.0 &   9.78\% \\
    \bottomrule
    \end{tabular}
    \caption{Ablation of annealing schedules for the soft-rounding temperature and the shape parameter of the Kumaraswamy noise on CLIC2020. Instead of annealing the soft-rounding temperature from 0.3 to 0.1 and the Kumaraswamy noise parameter from 2 to 1, we clamp them at the fixed value specified in the table. A higher BD-rate corresponds to worse RD performance relative to C3 with annealing.}
    \label{app:tab:ablation_annealing}
\end{table}

\subsection{Effect of Stage 2}
\label{app:sec:ablation:stage_2}

We also ablate the increase in performance that we can attribute to stage 2 of optimization, by comparing the default setting of \method (both stage 1 + 2) vs only having stage 1. We find that the BD-rate with respect to VTM for these two settings is $+1.39\%$ vs $+2.00\%$ on the CLIC2020 benchmark. The gain in BD-rate for stage 2 is indeed not as significant as was observed for COOL-CHICv2 \cite{leguay2023low}.

\subsection{BD-rate vs encoding iterations/time evaluation for CLIC and UVG}
\label{app:sec:ablation:encoding_iterations}
\begin{figure}[htb]
    \centering
    \includestandalone[mode=image]{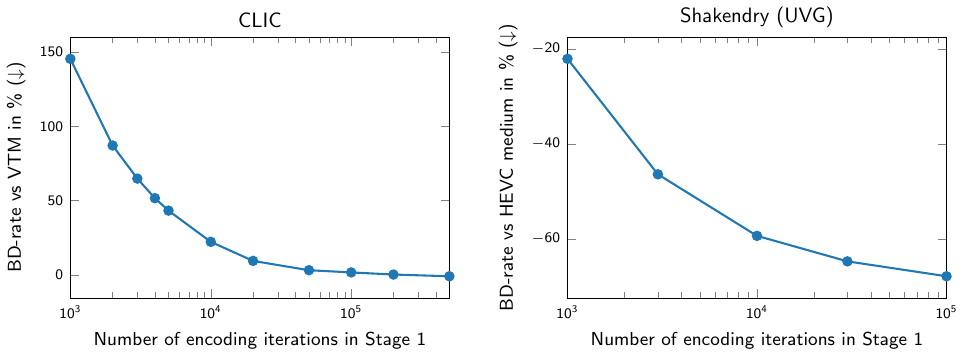}
    \caption{Ablation on the effect of using different number of encoding iterations in Stage 1 of the optimization. We evaluate the BD-rate (right) vs VTM (lower BD-rate values are better) on CLIC2020 and (left) vs HEVC on the Shakendry sequence of UVG. The number of encoding iterations in Stage 2 is determined adaptively but set to be at most $10\%$ of the number of iterations in Stage 1.}
    \label{app:fig:ablation:encoding_iterations}
\end{figure}

In \cref{app:fig:ablation:encoding_iterations} we show how the BD-rate of C3 changes as a function of the number of encoding iterations for both CLIC2020 and the Shakendry sequence of UVG. Note that with around $20-30$k iterations, we can approach the BD-rate of the default setting ($100$k iterations).

\subsection{Video ablations}
\label{app:sec:video_ablations}
\begin{figure}[ht]
    \centering
    \includestandalone[mode=image]{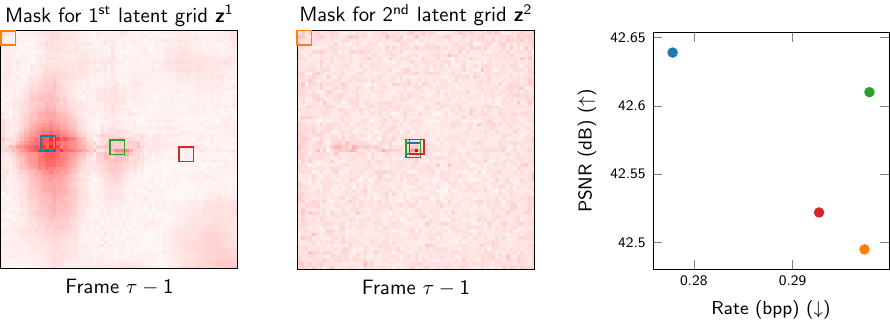}
    \caption{
    Comparison of (bpp, psnr) when training with different mask locations for the previous latent frame on the Jockey patch in \cref{app:fig:jockey}. The different locations are colour coded as follows:
    learned
        (\protect\tikz[baseline=-0.5ex]{\protect\draw[C0, thick] (0, 0) -- +(0.5, 0);}),
    top-left
        (\protect\tikz[baseline=-0.5ex]{\protect\draw[C1, thick] (0, 0) -- +(0.5, 0);}),
    center 
        (\protect\tikz[baseline=-0.5ex]{\protect\draw[C2, thick] (0, 0) -- +(0.5, 0);}),
    diametrically opposed to learned
        (\protect\tikz[baseline=-0.5ex]{\protect\draw[C3, thick] (0, 0) -- +(0.5, 0);}),
    }
    \label{app:fig:custom_mask_comparison}
\end{figure}

\begin{table}[h]
    \centering
    \begin{tabular}{lC}
    \toprule
    \textbf{Settings} & \textbf{BD-rate vs. HEVC (medium)} \\
    \midrule
    Single setting: \enticon{1} \synthicon{1} &  -7.67\% \\
    3 settings: single setting for \synthicon{1} but sweep \enticon{1}, \enticon{2}, \enticon{3} &  -21.44\% \\
    9 settings (default): sweep all combinations of \enticon{1}, \enticon{2}, \enticon{3} and \synthicon{1}, \synthicon{2}, \synthicon{3} & -28.89\%\\
    \bottomrule
    \end{tabular}
    \caption{UVG ablation for the $9$ hyperparameter settings used (cf. \cref{app:sec:hyperparameters}). Note that lower BD-rate means better RD performance.}
    \label{app:tab:ablation_uvg}
\end{table}

In \cref{app:fig:custom_mask_comparison}, we highlight the importance of learning the mask by comparing the rate and distortion values when using different custom mask locations for the previous latent grid. Among the four different choices of mask locations, we see that the learned mask (same as mask shown in \cref{app:fig:custom_mask}) achieves the best RD values.

In \cref{app:tab:ablation_uvg}, we show an ablation for how the BD-rate changes when we use a subset of the $9$ settings used for the video experiments on the UVG dataset (see \cref{app:tab:method:hyperparameters_video} for details on the $9$ settings). We see that varying the entropy settings and the synthesis settings are both important for improvements in performance.

\clearpage
\section{Additional visualizations}
\label{app:sec:additional_visualizations}

In this section, we show visualizations of C3 reconstructions and latents for both images and video. 

\begin{figure}[ht]
    \centering
    \includestandalone[mode=image]{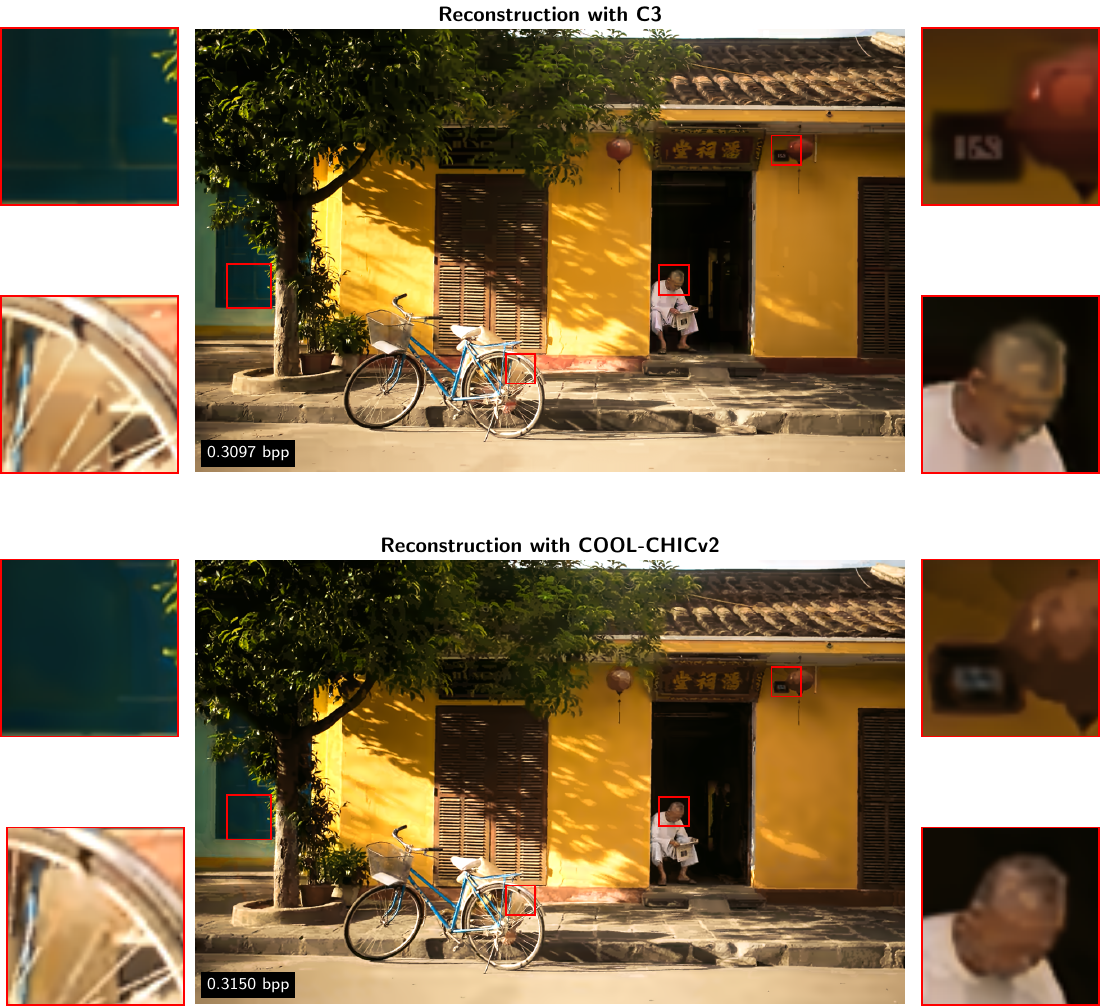}
    \caption{Qualitative comparison between \method (top) and COOL-CHIC v2 (bottom).
    The PSNR for \method is 30.28dB and the PSNR for COOL-CHIC v2 is 28.98dB.}
    \label{app:fig:c3_ccv2_comparison}
\end{figure}

\newcommand{\vidfigwidth}{0.43}

\begin{figure}[p]
    \small
    \centering
    \begin{subfigure}[b]{\vidfigwidth\textwidth}
        \centering
        \includegraphics[width=\textwidth]{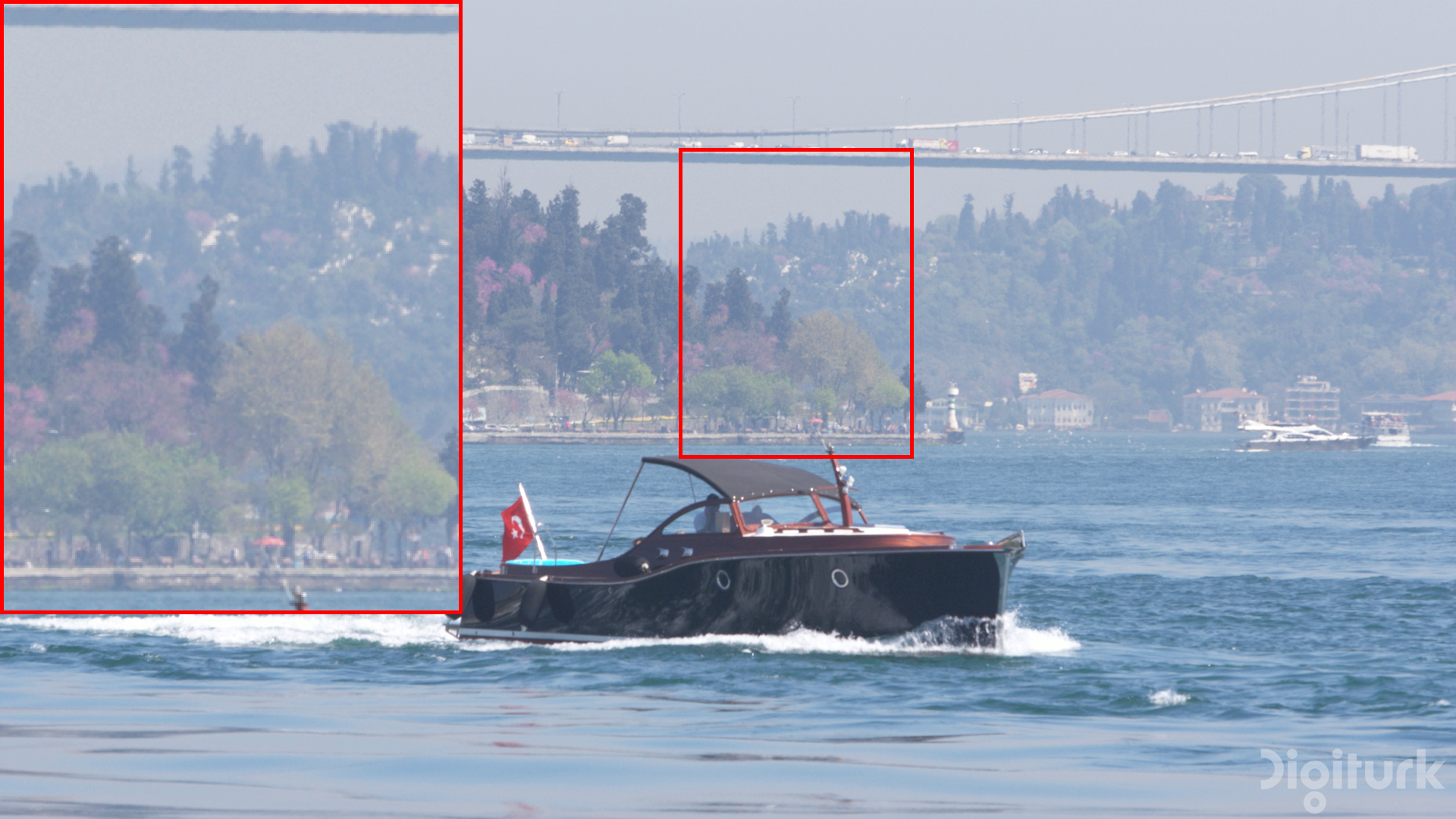}
        {Ground Truth \\ \hspace{1cm}}
    \end{subfigure}
    \begin{subfigure}[b]{\vidfigwidth\textwidth}
        \centering
        \includegraphics[width=\textwidth]{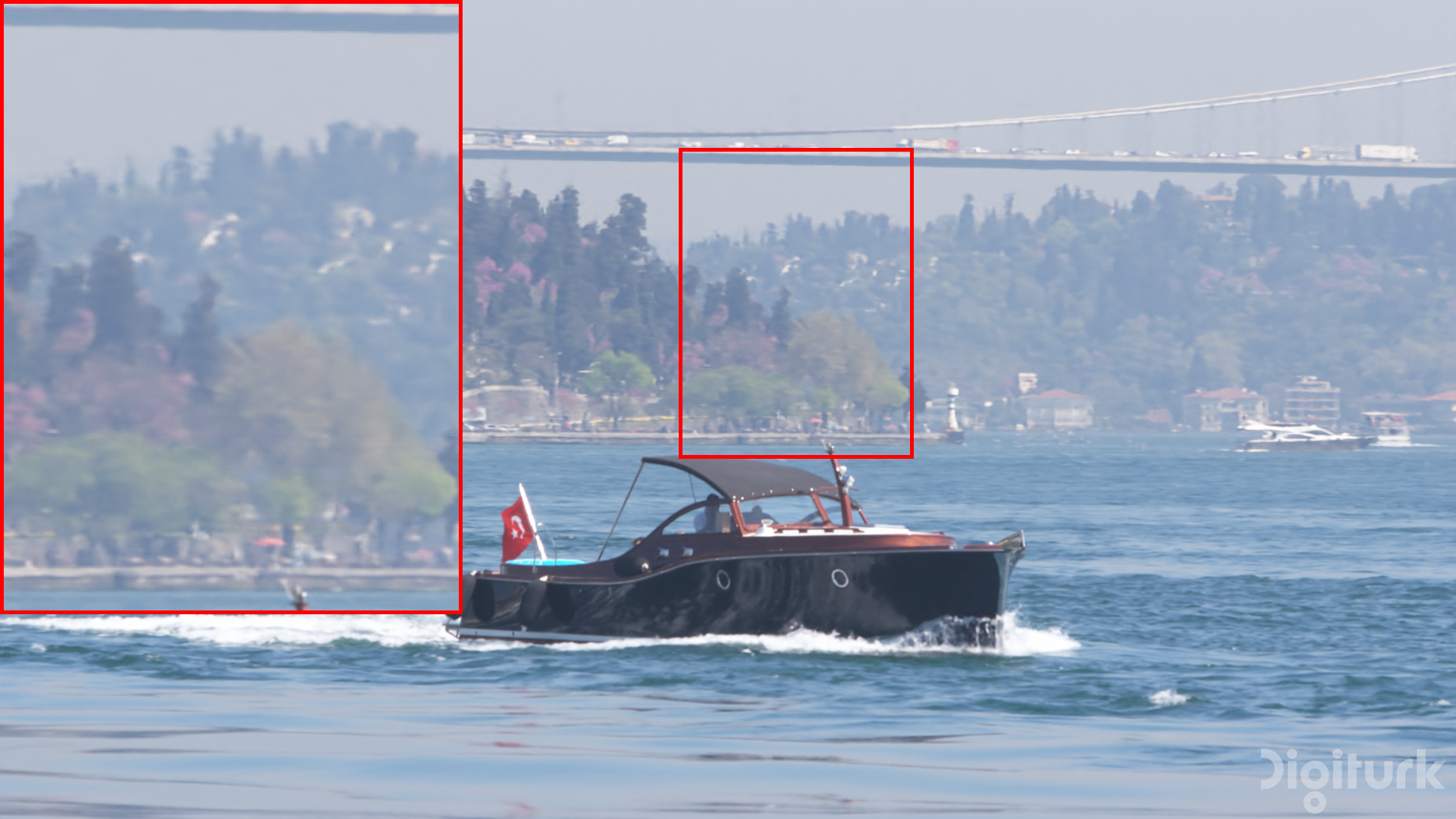}
        {HNeRV \cite{chen2023hnerv} \\ 39.1dB PSNR@0.101bpp}
    \end{subfigure}
    \begin{subfigure}[b]{\vidfigwidth\textwidth}
        \centering
        \includegraphics[width=\textwidth]{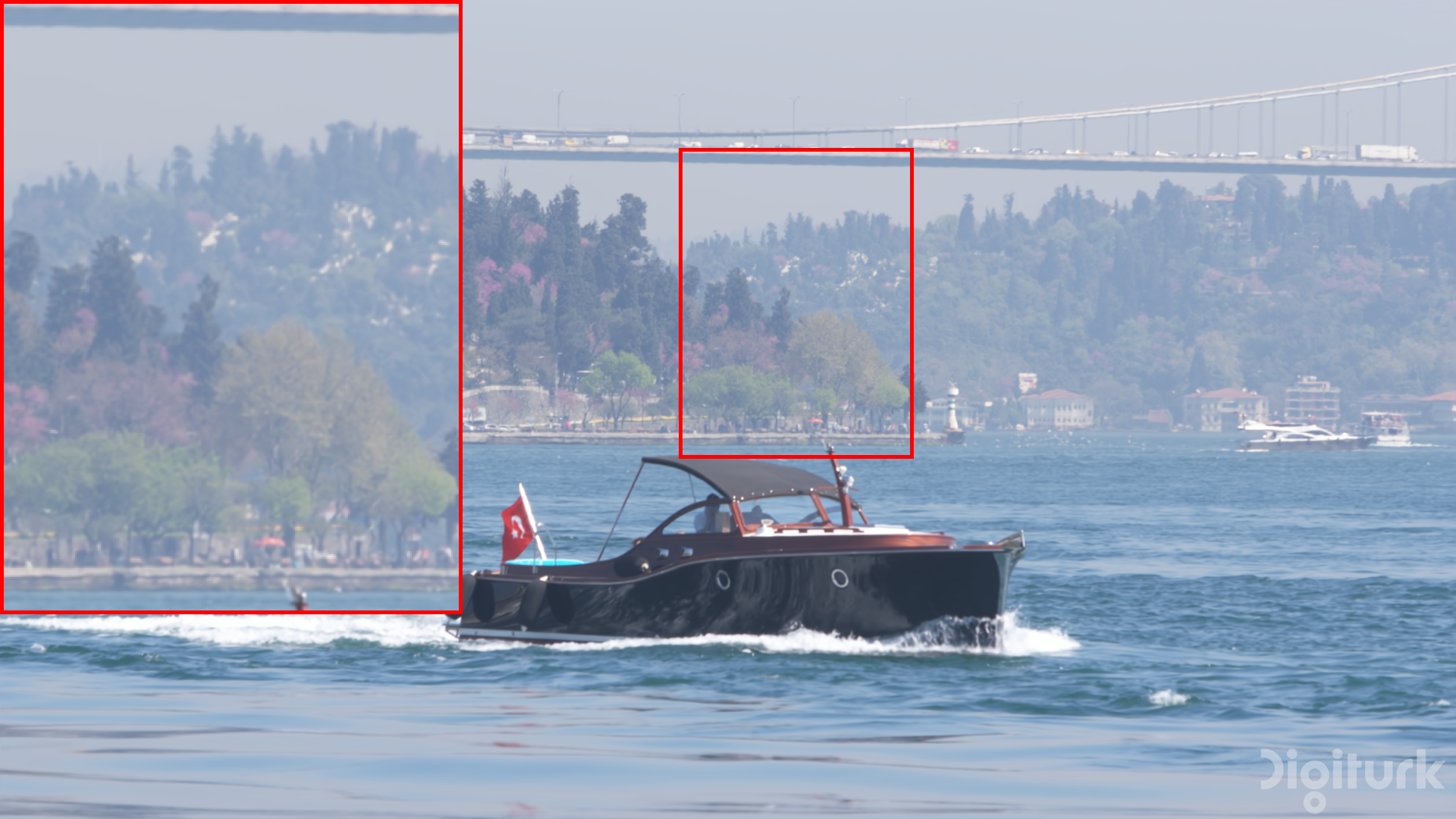}
        {HiNeRV \cite{kwan2023hinerv} \\ 41.1dB PSNR@0.051bpp}
    \end{subfigure}
    \begin{subfigure}[b]{\vidfigwidth\textwidth}
        \centering
        \includegraphics[width=\textwidth]{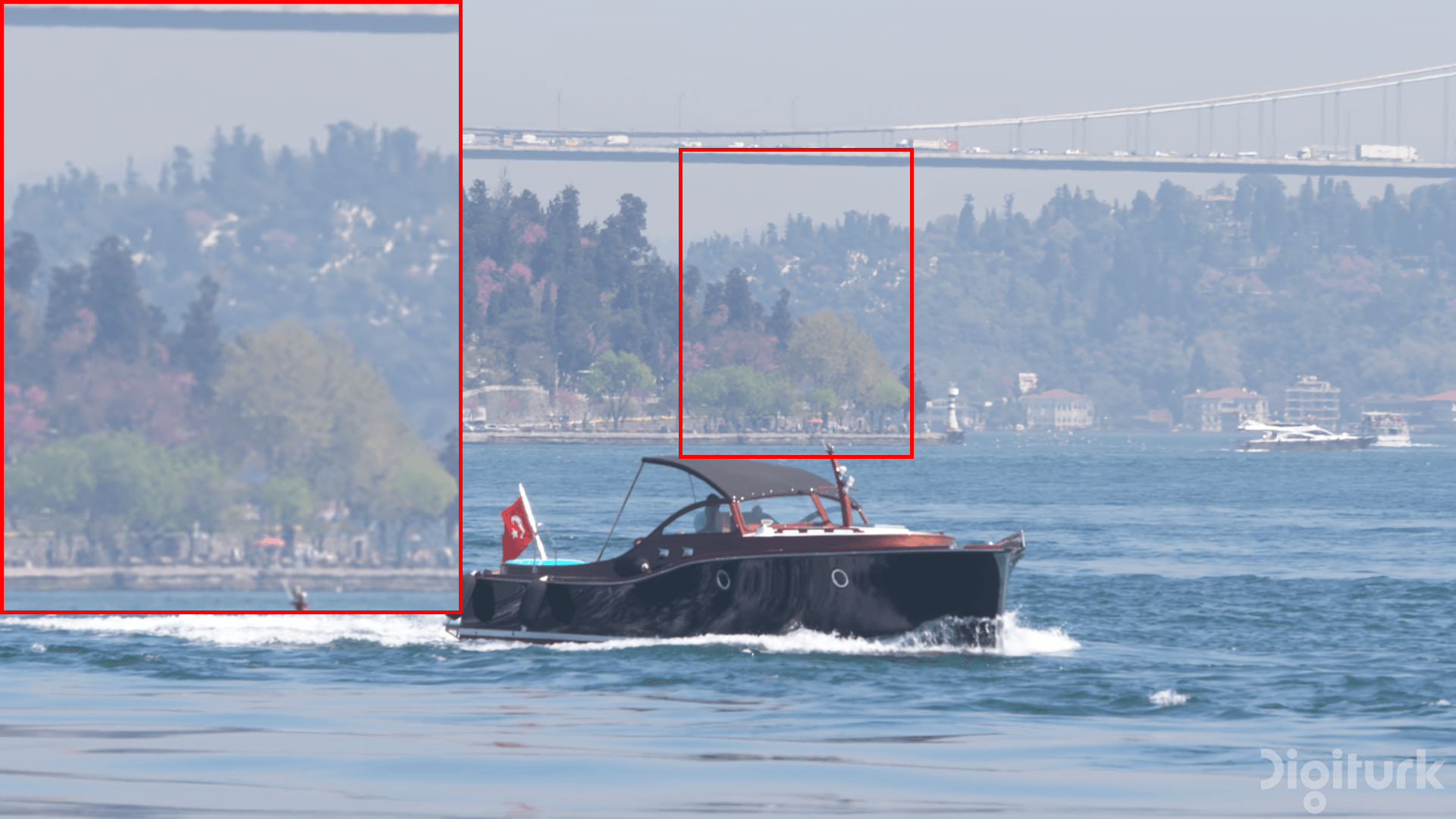}
        {C3 (ours) \\ 40.4dB PSNR@0.054bpp}
    \end{subfigure}
    \caption{Reconstructions of a frame of Bosphorus from the UVG dataset for various models. Adapted from Figure 9 of \citet{kwan2023hinerv}. 
    }
    \label{app:fig:uvg_rec}
\end{figure}

\begin{figure}[p]
    \small
    \centering
    \begin{tikzpicture}[every node/.style={inner sep=0pt}]
	\node[] (a) {\includegraphics[width=\vidfigwidth\textwidth]{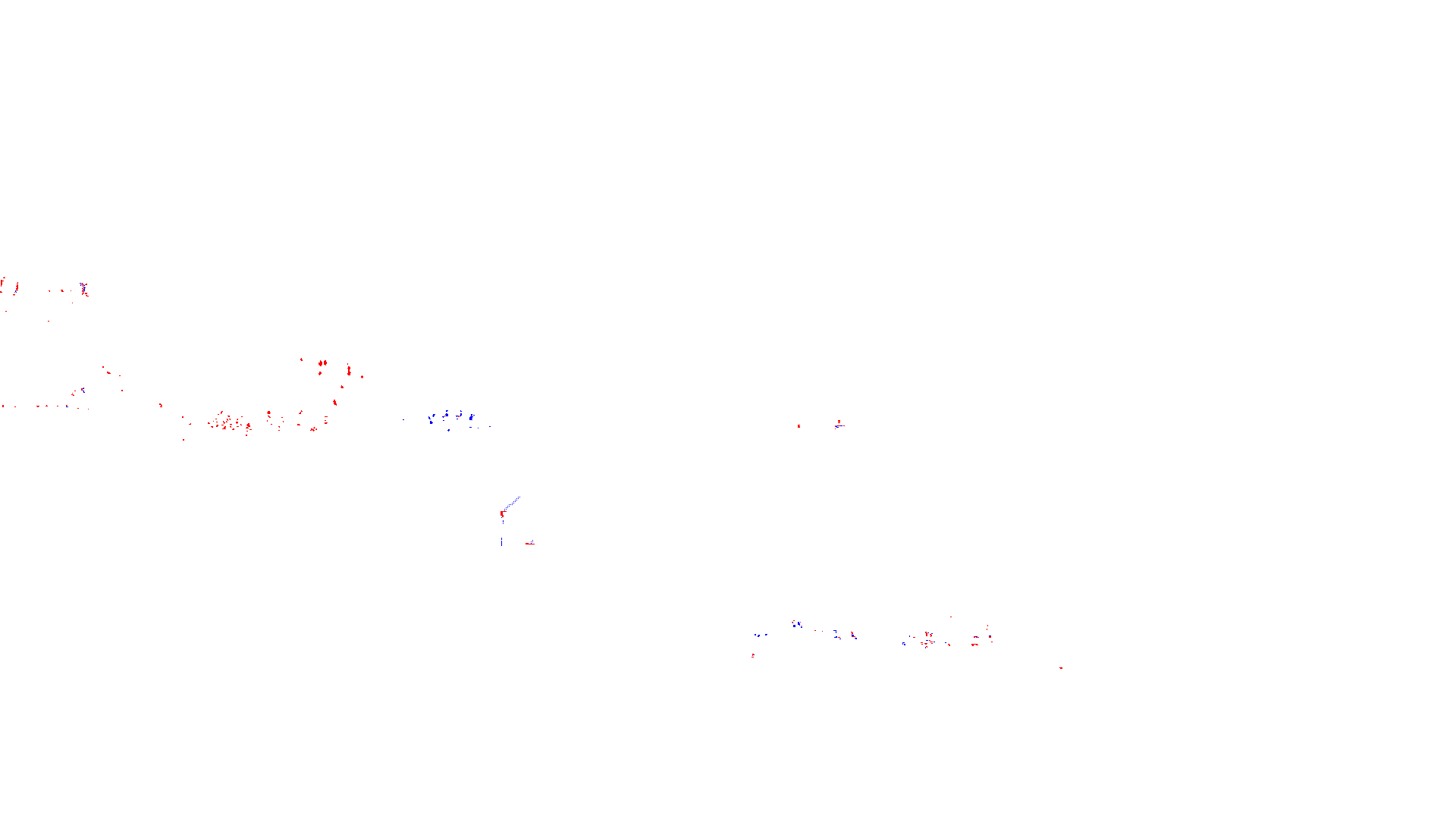}};
	\node[fit={(a)}, draw=darkgray, rectangle]{};
	\node[right=0.1 of a] (b) {\includegraphics[width=\vidfigwidth\textwidth]{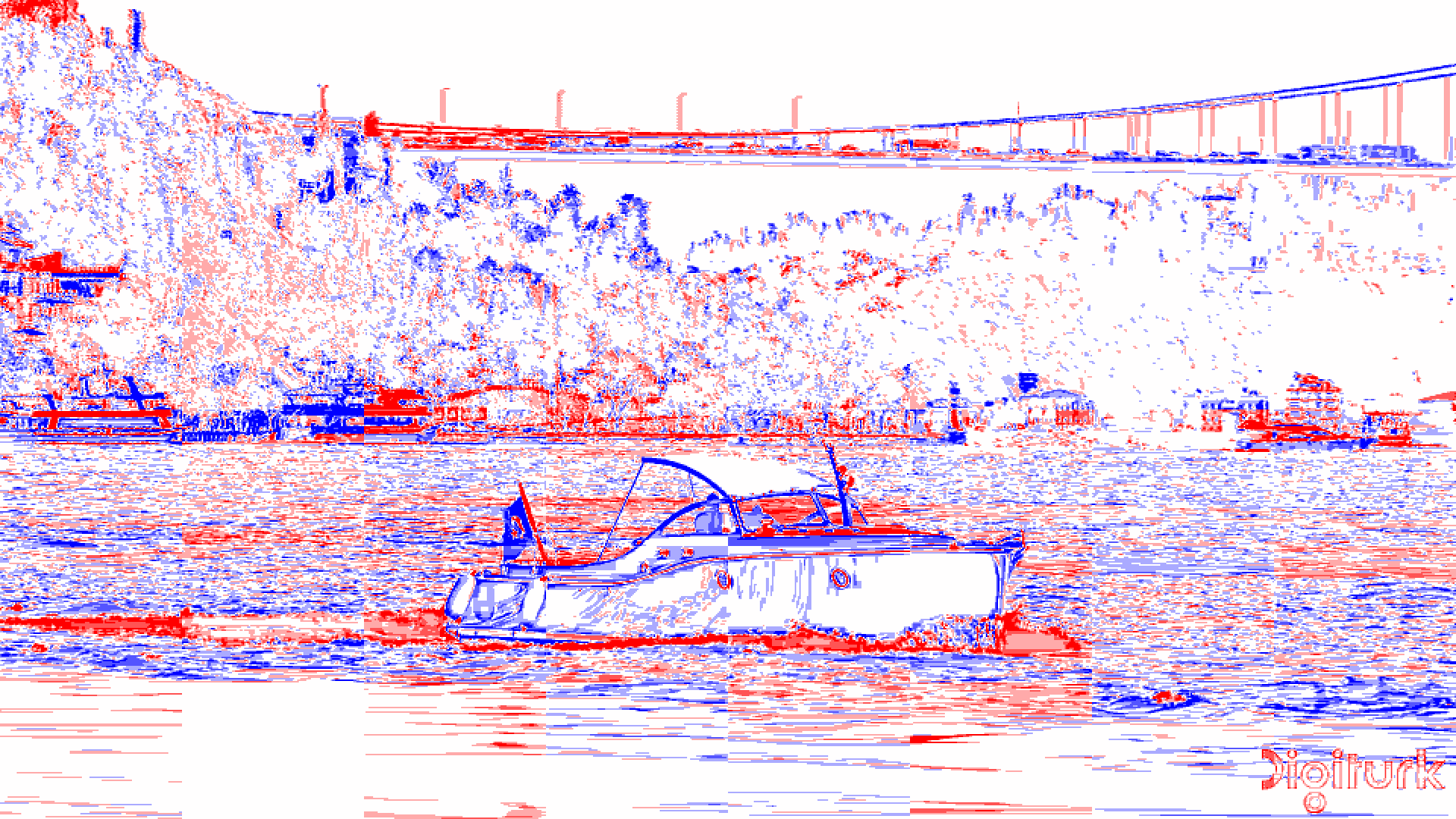}};
	\node[below=0.1 of a] (c) {\includegraphics[width=\vidfigwidth\textwidth]{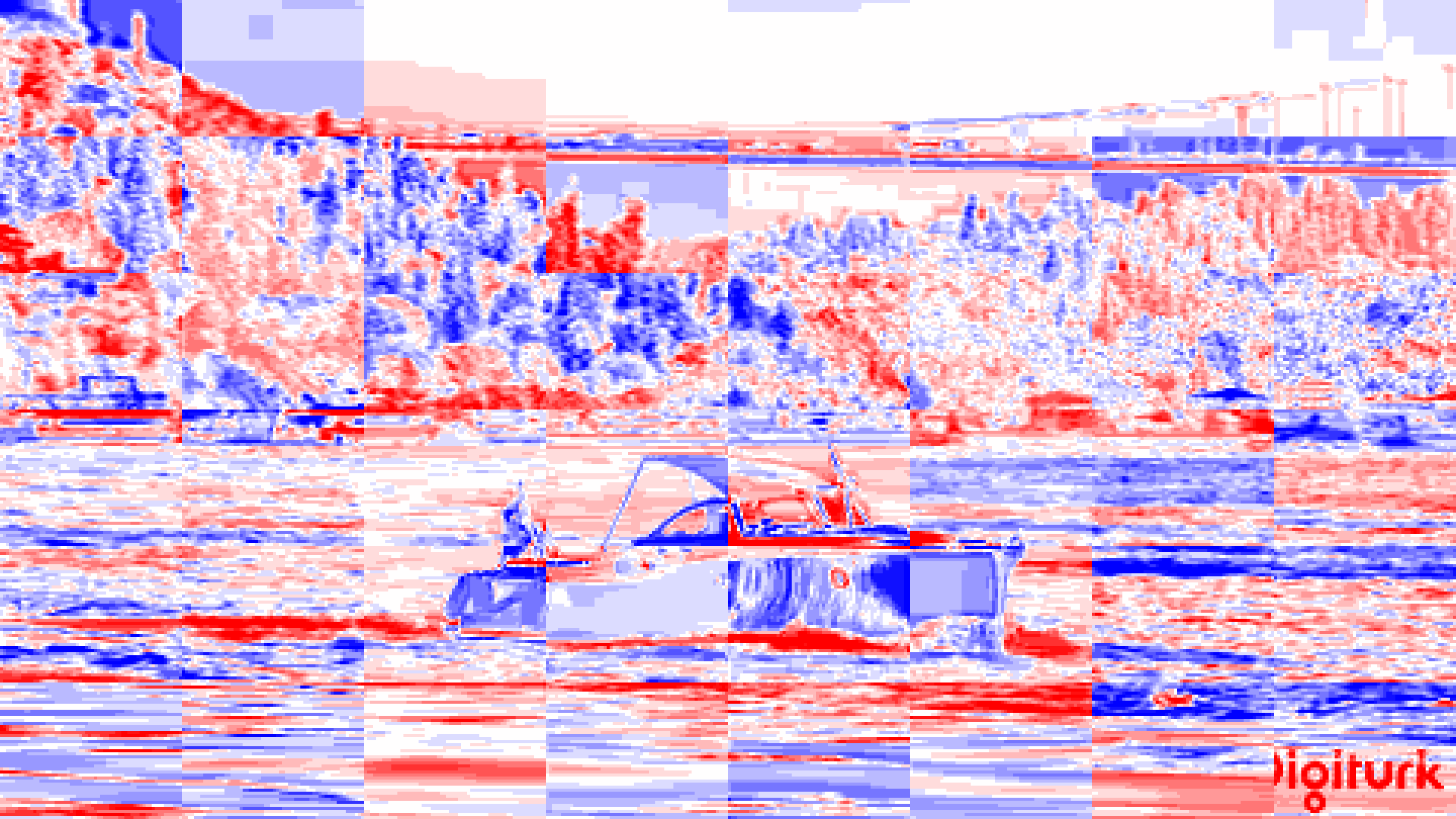}};
	\node[right=0.1 of c] (d) {\includegraphics[width=\vidfigwidth\textwidth]{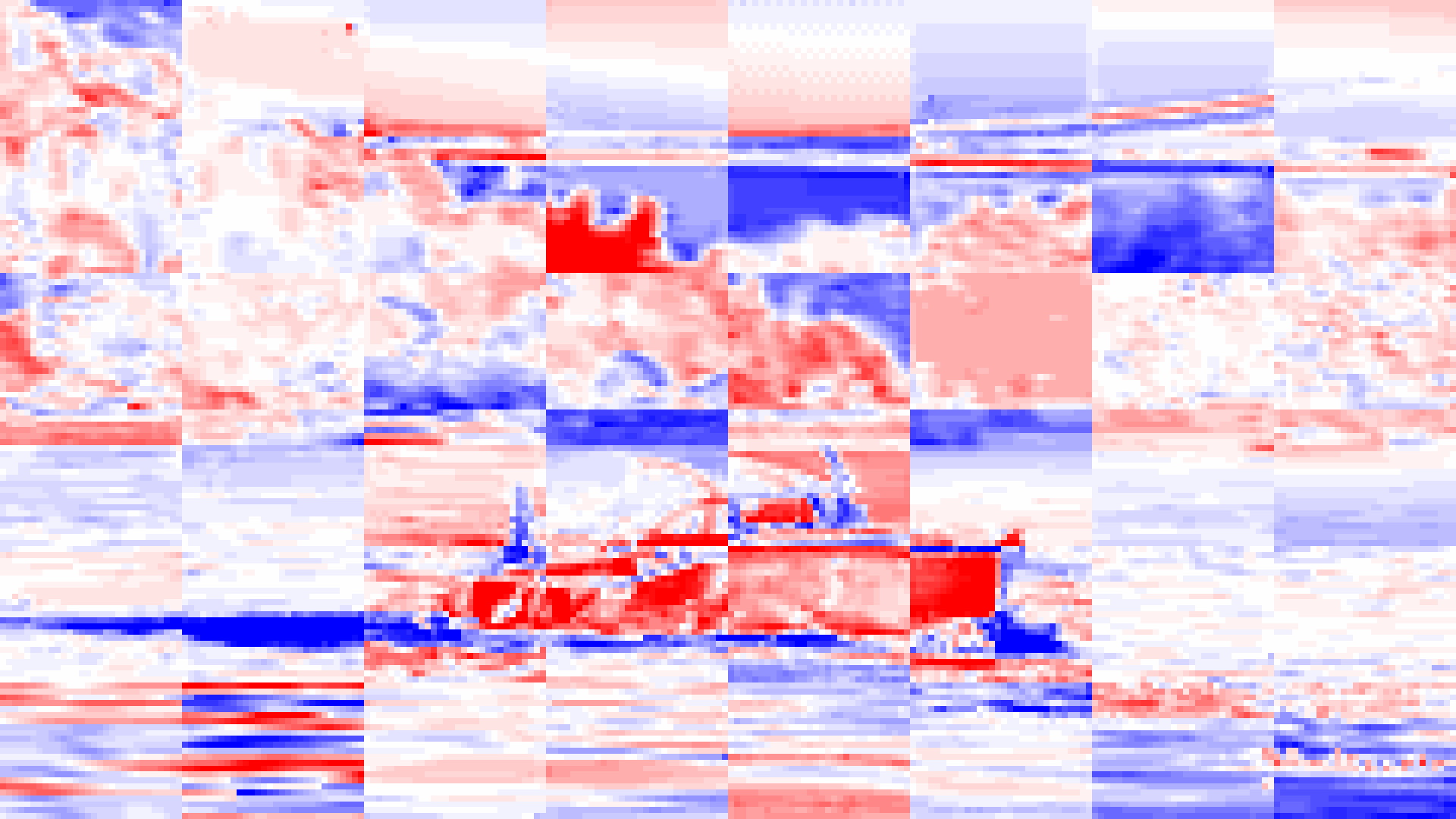}};
    \end{tikzpicture}
    \caption{\method latents of the first four grids corresponding to the frame in \cref{app:fig:uvg_rec}. Note that the highest resolution grid (top left) are mostly zeros, hence highly compressible. Also note that we see patch artifacts in the latents because each patch has been optimized independently, so we have different parameter values for the synthesis and entropy models of each patch. However we see in \cref{app:fig:uvg_rec} that these artifacts are not visible in reconstructions even for bpp values around 0.05.%
    }
    \label{app:fig:latents_vid}
\end{figure}
\clearpage 
\section{Raw values}

In this section we provide the raw values for the RD-curves of \method on all benchmarks.

\pgfplotstableset{%
columns/bpp/.style={fixed zerofill, precision=4, column type=c, column name=\textbf{Rate [bits per pixel]}},
columns/psnr/.style={fixed zerofill, precision=3, column type=c, column name=\textbf{PSNR [dB]},
},
empty cells with={--}, %
every head row/.style={before row=\toprule,after row=\midrule},
every last row/.style={after row=\bottomrule}
}

\begin{table}[htb]
\centering
\subcaptionbox{C3}{
    \begin{filecontents*}{rd_kodak_data_head.dat}
%% LaTeX2e file `rd_kodak_data_head.dat'
%% generated by the `filecontents' environment
%% from source `rd_kodak' on 2023/12/05.
%%
bpp, psnr
0.09379132045432925, 27.1242520014445
0.1351373130455613, 28.491956154505413
0.15518017454693714, 28.993290583292644
0.2041274414708217, 30.05834722518921
0.2333232325812181, 30.601855913798016
0.2784953669955333, 31.336687008539837
0.35267731361091137, 32.43605081240336
0.5181328505277634, 34.3912710348765
0.5808481611311436, 35.039207776387535
0.7309025563299656, 36.39331324895223
0.8162287150820097, 37.00826279322306
0.9216941706836224, 37.8352115948995
1.0977161079645157, 38.895706494649254
1.4105306218067806, 40.86893860499064
\end{filecontents*}
\pgfplotstabletypeset[col sep=comma, columns={bpp, psnr}]{rd_kodak_data_head.dat}} \hspace{1cm}
\subcaptionbox{C3 \emph{adaptive}}{
    \begin{filecontents*}{rd_kodak_data_head-best-per-image.dat}
%% LaTeX2e file `rd_kodak_data_head-best-per-image.dat'
%% generated by the `filecontents' environment
%% from source `rd_kodak' on 2023/12/05.
%%
bpp, psnr
0.07914485937605302, 27.049038410186768
0.12280272568265597, 28.38341196378072
0.14266458339989185, 28.859963019688923
0.19369436024377742, 29.922741810480755
0.2241452575350801, 30.472349564234417
0.2708829619611303, 31.24034849802653
0.349332923690478, 32.389469067255654
0.5175360428790251, 34.403931538263954
0.5804239014784495, 35.05256875356039
0.7295577526092529, 36.40783214569092
0.8067246576150259, 37.05852699279785
0.9135203249752522, 37.89463567733765
1.0812142690022786, 39.03287013371786
1.4080520172913868, 40.91846783955892
\end{filecontents*}
\pgfplotstabletypeset[col sep=comma, columns={bpp, psnr}]{rd_kodak_data_head-best-per-image.dat}}
\caption{Raw values of our proposed method, \method, on the Kodak image benchmark.}
\label{app:tab:raw_values_kodak}
\end{table}

\begin{table}[htb]
\centering
\subcaptionbox{C3}{
    \begin{filecontents*}{rd_clic_data_head.dat}
%% LaTeX2e file `rd_clic_data_head.dat'
%% generated by the `filecontents' environment
%% from source `rd_clic' on 2023/12/05.
%%
bpp, psnr
0.06225230403971381, 29.158539423128452
0.09255141524098269, 30.541110387662563
0.10576727668323167, 31.0040992178568
0.14050452093162188, 32.035784000303686
0.16096663711274542, 32.50190418522532
0.19082382984641122, 33.17557734977908
0.24083542751102913, 34.079383291849275
0.351732343253566, 35.71583850209306
0.3951945726464434, 36.24175634616759
0.49943820659707233, 37.33614302844536
0.5648835810219369, 37.803539183081654
0.650084137553122, 38.470385109506005
0.7841241294290962, 39.445163819848034
1.0723320667336627, 40.95862235092535
\end{filecontents*}
\pgfplotstabletypeset[col sep=comma, columns={bpp, psnr}]{rd_clic_data_head.dat}}
    \hspace{1cm}
\subcaptionbox{C3 \emph{adaptive}}{
    \begin{filecontents*}{rd_clic_data_head-best-per-image.dat}
%% LaTeX2e file `rd_clic_data_head-best-per-image.dat'
%% generated by the `filecontents' environment
%% from source `rd_clic' on 2023/12/05.
%%
bpp, psnr
0.05416938280913888, 29.089758337997807
0.08611200604496932, 30.531509166810572
0.09947567428575783, 31.01140878258682
0.13428194498325266, 32.018704205024534
0.15572169741115918, 32.5368369497904
0.18597072889892066, 33.18091597208163
0.23686866235078835, 34.10931619783727
0.3469772873128333, 35.69953760286657
0.39149349001122685, 36.24025112245141
0.497362180453975, 37.32819301326101
0.5560192167758942, 37.82749985485542
0.6443476524294877, 38.52037299551615
0.7797551020616438, 39.460437867699596
1.0665377756444419, 41.00582569401438
\end{filecontents*}
\pgfplotstabletypeset[col sep=comma, columns={bpp, psnr}]{rd_clic_data_head-best-per-image.dat}}
\caption{Raw values of our proposed method, \method, on the CLIC2020 professional validation dataset split image benchmark.}
\label{app:tab:raw_values_clic}
\end{table}

\begin{table}[htb]
\centering
\subcaptionbox{C3 on all UVG videos}{
    \begin{filecontents*}{rd_uvg_data_head.dat}
%% LaTeX2e file `rd_uvg_data_head.dat'
%% generated by the `filecontents' environment
%% from source `rd_uvg_data' on 2023/12/05.
%%
bpp, psnr
0.01594833859347827, 32.925598689488005
0.02268396719193247, 33.96687752859933
0.033139908094155894, 35.02668489728655
0.05459793865953716, 36.327970777239116
0.08459428009726773, 37.44999258858817
0.12762070219697697, 38.36397443498884
0.25398065725859764, 39.704908643450054
0.44254081146819296, 40.794677189418245
\end{filecontents*}
\pgfplotstabletypeset[col sep=comma, columns={bpp, psnr}]{rd_uvg_data_head.dat}} \\[1em]
\subcaptionbox{C3 on Beauty}{
    \begin{filecontents*}{rd_uvg_beauty_data_head_full.dat}
%% LaTeX2e file `rd_uvg_beauty_data_head_full.dat'
%% generated by the `filecontents' environment
%% from source `rd_uvg_individual_videos' on 2023/12/05.
%%
bpp, psnr
0.010050684601689378, 32.97476577758789
0.012360346799444718, 33.41041946411133
0.0160525635680339, 33.735130310058594
0.024176362251940493, 34.06508255004883
0.04204665634121436, 34.28904342651367
0.10816917172245061, 34.936195373535156
0.44460512317406636, 36.891178131103516
0.9687660108630856, 38.808494567871094
\end{filecontents*}
\pgfplotstabletypeset[col sep=comma, columns={bpp, psnr}]{rd_uvg_beauty_data_head_full.dat}}
\subcaptionbox{C3 on Bosphorus}{
    \begin{filecontents*}{rd_uvg_bosphorus_data_head_full.dat}
%% LaTeX2e file `rd_uvg_bosphorus_data_head_full.dat'
%% generated by the `filecontents' environment
%% from source `rd_uvg_individual_videos' on 2023/12/05.
%%
bpp, psnr
0.010261636331658034, 35.56731033325195
0.014262584608028797, 36.722042083740234
0.02033674122018662, 37.918617248535156
0.032516769657377154, 39.32243347167969
0.053004607123633224, 40.383018493652344
0.07365888472801695, 41.23855972290039
0.13820952924628122, 42.4293098449707
0.20775837127002889, 43.286441802978516
\end{filecontents*}
\pgfplotstabletypeset[col sep=comma, columns={bpp, psnr}]{rd_uvg_bosphorus_data_head_full.dat}}
\subcaptionbox{C3 on Honeybee}{
    \begin{filecontents*}{rd_uvg_honeybee_data_head_full.dat}
%% LaTeX2e file `rd_uvg_honeybee_data_head_full.dat'
%% generated by the `filecontents' environment
%% from source `rd_uvg_individual_videos' on 2023/12/05.
%%
bpp, psnr
0.008884227795836827, 35.78207015991211
0.010703726678912062, 36.51614761352539
0.013055100256072668, 37.125831604003906
0.017905752916703932, 37.793296813964844
0.03134384720081774, 38.39179229736328
0.041387882040968785, 38.73643112182617
0.08882348972450321, 39.1976203918457
0.19873188129470995, 39.88926315307617
\end{filecontents*}
\pgfplotstabletypeset[col sep=comma, columns={bpp, psnr}]{rd_uvg_honeybee_data_head_full.dat}} \\[0.5em]
\subcaptionbox{C3 on Jockey}{
    \begin{filecontents*}{rd_uvg_jockey_data_head_full.dat}
%% LaTeX2e file `rd_uvg_jockey_data_head_full.dat'
%% generated by the `filecontents' environment
%% from source `rd_uvg_individual_videos' on 2023/12/05.
%%
bpp, psnr
0.015441530968625253, 31.970964431762695
0.021698799939864937, 33.01723098754883
0.032144206794328056, 34.09822082519531
0.055076323231332935, 35.462181091308594
0.08609376446499178, 36.860538482666016
0.12893884191289545, 37.86589050292969
0.22348828334361315, 39.1264762878418
0.3798738466886183, 40.033348083496094
\end{filecontents*}
\pgfplotstabletypeset[col sep=comma, columns={bpp, psnr}]{rd_uvg_jockey_data_head_full.dat}}
\subcaptionbox{C3 on Readysetgo}{
    \begin{filecontents*}{rd_uvg_readysetgo_data_head_full.dat}
%% LaTeX2e file `rd_uvg_readysetgo_data_head_full.dat'
%% generated by the `filecontents' environment
%% from source `rd_uvg_individual_videos' on 2023/12/05.
%%
bpp, psnr
0.03440686540610235, 29.709619522094727
0.05015435066040178, 31.278114318847656
0.07315845542567938, 32.8843994140625
0.11821751421909237, 34.93806076049805
0.1684759724206136, 36.799190521240234
0.23755911035696045, 38.19867706298828
0.364017710032446, 40.0916748046875
0.523060373775661, 41.219112396240234
\end{filecontents*}
\pgfplotstabletypeset[col sep=comma, columns={bpp, psnr}]{rd_uvg_readysetgo_data_head_full.dat}}
\subcaptionbox{C3 on Shakendry}{
    \begin{filecontents*}{rd_uvg_shakendry_data_head_full.dat}
%% LaTeX2e file `rd_uvg_shakendry_data_head_full.dat'
%% generated by the `filecontents' environment
%% from source `rd_uvg_individual_videos' on 2023/12/05.
%%
bpp, psnr
0.01135890802349119, 32.94499206542969
0.016844887878202524, 33.900936126708984
0.026078339738887735, 34.94967269897461
0.045114195866820715, 36.1951904296875
0.075923013399976, 37.218631744384766
0.11065860791131854, 37.9559211730957
0.20995707732314864, 38.823490142822266
0.38954471551502745, 39.84563446044922
\end{filecontents*}
\pgfplotstabletypeset[col sep=comma, columns={bpp, psnr}]{rd_uvg_shakendry_data_head_full.dat}}\\[0.5em]
\subcaptionbox{C3 on Yachtride}{
    \begin{filecontents*}{rd_uvg_yachtride_data_head_full.dat}
%% LaTeX2e file `rd_uvg_yachtride_data_head_full.dat'
%% generated by the `filecontents' environment
%% from source `rd_uvg_individual_videos' on 2023/12/05.
%%
bpp, psnr
0.02123451702694486, 31.529468536376953
0.03276307377867246, 32.92325210571289
0.051153949655902885, 34.47492218017578
0.08917865247349255, 36.51955032348633
0.13527209972962737, 38.207733154296875
0.192972416706228, 39.616146087646484
0.30876338796612496, 41.374610900878906
0.4300504808702196, 42.480445861816406
\end{filecontents*}
\pgfplotstabletypeset[col sep=comma, columns={bpp, psnr}]{rd_uvg_yachtride_data_head_full.dat}}
\caption{Raw values of our proposed method, \method, UVG video benchmark; average rate and PSNR over the seven 1080p videos.}
\label{app:tab:raw_values_uvg}
\end{table}

\end{document}